# A Review of Personality in Human–Robot Interactions

Foundations and Trends® in Information Systems

*Lionel P. Robert Jr.
Associate Professor
School of Information
Robotics Institute
University of Michigan
Ann Arbor, MI 48109-1285

**Rasha Alahmad
School of Information
University of Michigan
Ann Arbor, MI 48109-1285

Connor Esterwood
School of Information
University of Michigan
Ann Arbor, MI 48109-1285

Sangmi Kim
School of Information
University of Michigan
Ann Arbor, MI 48109-1285

Sangseok You
Assistant Professor
HEC Paris
1 Rue de la Liberation,
Jouy-en-Josas, France

Qiaoning Zhang
School of Information
University of Michigan
Ann Arbor, MI 48109-1285

## Abstract

Personality has been identified as a vital factor in understanding the quality of human–robot interactions. Despite this the research in this area remains fragmented and lacks a coherent framework. This makes it difficult to understand what we know and identify what we do not. As a result, our knowledge of personality in human–robot interactions has not kept pace with the deployment of robots in organizations or in our broader society. To address this shortcoming, this paper reviews 83 articles and 84 separate studies to assess the current state of human–robot personality research. This review: (1) highlights major thematic research areas, (2) identifies gaps in the literature, (3) derives and presents major conclusions from the literature and (4) offers guidance for future research.

*Corresponding author
**The following names are listed in alphabetical order.

# Table of Contents

# Chapter 1

# Introduction

Robots -- technologies that can sense, reason and respond to their environments through embodied actions -- are being used in new domains to both replace and complement humans (You and Robert 2018; You, Kim, Lee, Kamat, Robert,2018). This means robots are interacting with an organization's employees and in some cases directly interacting with their customers. The need for robots to directly interact with humans has led many researchers to identify factors that promote human–robot interaction. Personality has been identified as a vital factor in understanding the nature and quality of human–robot interactions (Gockley and Matarić 2006; Goetz and Kiesler 2002; Robert 2018; Syrdal, Dautenhahn, Woods, Walters, and Koay 2007). What is personality? Personality comprises someone's past behaviors, cognitions and emotions derived from both biological and social factors (Hall and Lindzey 1957). Why would scholars turn to personality to understand human–robot interaction? To answer these questions, this volume turns to the organizational behavior and social psychology literature on personality. However, given the paper's focus on personality as it relates to human–robot interaction, the discussion will be brief.

Theories of personality assert that individual human traits can be used to predict human emotions, cognitions and behaviors (Peeters et al. 2006). "Personality traits" is a label to describe a specific set of characteristics that are believed to be the best predictors of an individual's behavior (Tasa, Sears and Schat 2011). Personality is now considered a core construct in understanding human behavior over and above many other factors (Li, Barrick, Zimmerman and Chiaburu 2014). More important, personality explains the way people respond to others in social settings (Thoresen, Kaplan, Barsky, Warren, de Chermont et al. 2003). This is why personality influences the quality of interactions between individuals (Driskell, Goodwin, Salas, and O'Shea 2006; Peeters et al. 2006). The literature on personality is rich in theory and spans disciplines such as sociology, psychology, and political science as well as organizational behavior.

Although there are many types of personality traits, the Big Five are held in particularly high regard. The Big Five personality traits are the most widely used personality traits (Li et al. 2014). The acronym OCEAN, representing openness to experience, conscientiousness, extraversion, agreeableness and neuroticism, is often used to represent the five personality traits. Openness to experience represents the degree to which someone is imaginative, curious, and broadminded (McCrae and Costa 1997). Conscientiousness reflects the extent that someone is careful, deliberative and self-aware of their actions (Tasa et al. 2011). Extraversion is the extent to which an individual is assertive, outgoing, talkative, and sociable (Rhee, Parent and Basu 2013). Introversion is the degree to which someone enjoys being alone and is the opposite of extraversion (Driskell et al. 2006). Agreeableness reflects the extent to which someone is cooperative and friendly (Peeters et al. 2006). Neuroticism can be viewed as the degree to which someone is easily angered, not well-adjusted, insecure, and lacks self-confidence (Driskell et al. 2006). Neuroticism is often viewed as the opposite of emotional stability, which is the degree to which someone is calm, well-adjusted, secure, and self-confident (Peeters et al. 2006). The Big Five are not only the most popular set of personality traits in social sciences, but, as we demonstrate here, they are also the most popular traits used in the study of human–robot interaction (Robert 2018).

Despite the importance of personality in the HRI literature, the research remains fragmented and lacks a coherent framework. This makes it difficult to understand what we know and identify what we do not. As a result, our knowledge of personality in human–robot interactions has not kept pace with the deployment of robots in organizations or in our broader society. As robots become increasingly vital to our society, there is a need to better comprehend factors such as personality that facilitate better human–robot interaction (HRI).

To address this shortcoming, this paper reviews the literature on personality and embodied physical action (EPA) robots. We focused on EPA robots because their physical embodiment invokes strong emotional reactions that can lead individuals to project personalities onto them (Robert 2018; You and Robert 2018). Therefore, issues related to personality are likely to be more central to human–robot interaction with regard to EPA robots. This paper investigates the current state of human–robot personality research, discusses the unique role of personality in human–robot research, and offers guidance for future research.

This review offers several contributions to the literature. First, it presents a conceptual integrated model of the literature on personality in human–robot literature. In doing so, this paper helps to organize the literature on personality in human–robot literature. Two, it highlights four thrust areas in the literature. These thrust areas include: (1) Human Personality and HRI, (2) Robot Personality and HRI, (3) Robot Personality and HRI, and (4) Factors Impacting Robot Personality. Three, it derives and presents major insights from the literature. Finally, it identifies gaps in the literature that need to be addressed.

The paper is organized as follows. Next, in Chapter 2, we present the relevant literature including the inclusion and exclusion criteria for articles. This includes a brief discussion of the publication venues, personality measures, and outcome measures in the literature. Then, in Chapter 3 we present and discuss Thrust Area 1: Human Personality and HRI. In Chapters 4, 5, and 6, a similar discussion takes place for Thrust Area 2: Robot Personality and HRI, Thrust Area 3: Robot Personality and HRI, and Thrust Area 4: Factors Impacting Robot Personality, respectively. Chapter 7 follows with a discussion on the way forward, focusing on the opportunities for personality research in human–robot interaction. Finally, the paper is concluded in Chapter 8.

In summary, robots are being used to both replace and complement humans across many settings. Personality has been identified as a vital factor in the promotion of human–robot interaction. Unfortunately, the HRI personality literature lacks a coherent framework, making it difficult to comprehend how personality can facilitate better human–robot interaction (HRI). To address this problem, we review the current state of human–robot personality research in hopes of providing guidance for future research.

# Chapter 2

# Literature Search

The literature review employed several search engines: Google Scholar, ACM Digital Library, Scopus, PsycINFO, and IEEE Xplore. The search was conducted in December 2017.

## Section 2.1. Study Selection Process

*Search Terms*. There were three main search terms. The first search term included the words "human robot interaction and personality," the second included "robot personality," and the third search term included "HRI and personality." We used these search terms across all search engines. In all cases the search terms yielded a return of thousands of potential articles presented in order of relevance to the topic. The literature search involved going through each article based on the initial inclusion criteria until the results page yielded no more relevant articles. Beyond this point, the articles only included the terms robot or personality but not both. These articles either discussed human–robot interaction in absence of personality or human personality in absence of the robot.

*Initial Inclusion Criteria*. The initial search yielded 220 unique articles. Studies were initially included if they explicitly mentioned both the terms "robot" and "personality" and were published in English-language journals/conferences.

*Final Inclusion Criteria*. Studies were included if they (1) were empirical studies using EPA robots, (2) measured human or robot personality and conducted a study involving humans interacting with EPA robots.

*Exclusion Criteria*. Studies were excluded if they (1) focused on embodied virtual action (EVA) (i.e. virtual agents), (2) focused on tele-presence robots, (3) focused only on manipulating robot personality without examining its impact on humans or (4) focused only on Negative Attitudes toward Robots (NARS) as the personality trait of interest. NARS is normally used as control variable in many studies (see You and Robot, 2018). It would be of no surprise that it would be negatively correlated to any measure of human–robot interaction.

After screening the initial 220 articles against the final inclusion and exclusion criteria, we had 83 empirical articles with 84 separate studies on the topic of human–robot interaction of EPA robots. In this process we eliminated 15 non-empirical articles on the topic of human–robot interaction EPA robots. The 15 non-empirical articles were primarily technical descriptions of various personality-based systems. There were 119 articles on the topics of embodied virtual action (EVA) (i.e. virtual agents) or tele-presence robots, or that used NARS as the only personality trait of interest.

## Section 2.2. Publication Venues

The publication venues of the included articles were as follows: 73.6% were published in conferences while 26.4% were published in journals. The Human–Robot Interaction (HRI) Conference accounted for the most included articles, with 26.3% of all the articles and 35.7% of the conference publications. This was followed by IEEE International Symposium on Robot and Human Interactive Communication (ROMAN), which accounted for 10.5% of all publications and 14.2% of all conferences. Publication dates ranged from 2005 to 2017 (Figure 2.1).

## Section 2.3. Publications by Year

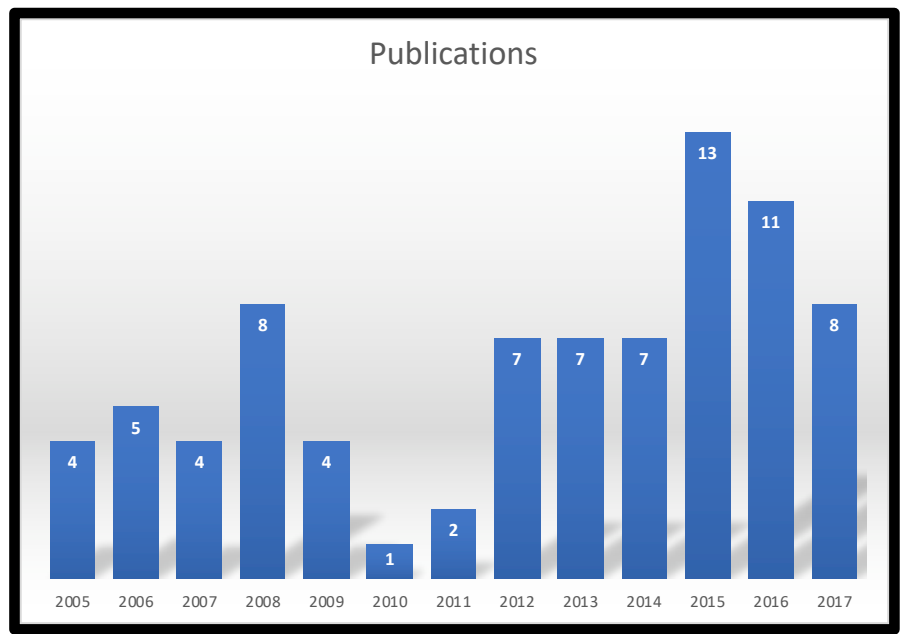

Figure 2.1 Publications by Year

## Section 2.4. Personality Measures

The Big Five personality characteristics were by far the most widely used measures. More than 89% of the articles employed some measure of one or more of them. Measures of extraversion/introversion were the most popular. They were the only Big Five personality measures included in every study employing some measure of the Big Five.

## Section 2.5. Outcome Measures

Measures used to assess the quality of human–robot interactions varied. However, measures of effectiveness were the most popular (31.5%). These included feelings of attachment and liking toward the robot. Trust or confidence in the robot was next (26.3%). Several other outcomes were more equally represented, such as the distance one is comfortable interacting with robots (15.7%), acceptance (15.7%), preference for a particular robot (15.7%), and compliance with a robot's advice or suggestions (15.7%). Most studies had more than one outcome variable.

## Section 2.6. Sample

***Number of Participants.*** Nearly all studies reported the number of participants. The average sample size was 60 but the standard deviation was 74.77, indicating that sample size per study varied greatly. Further analysis revealed that the median was 36 and the mode was 32.

***Age.*** In the 84 studies, 58 studies directly reported the average age. For those studies, the reported average age of participants was 31.73 years, with a standard deviations of 6.67 years. However, 26 studies did not directly report the average age: 15 reported an average range (e.g., 21–27 years) and 11 did not provide any information regarding the participants' average age.

***Gender Diversity.*** In all, 63 studies reported the number of men and women participants. Among studies that reported the number of men and women participants, 45.8% of the participants were women. Twenty-one studies did not provide the number of men and women participants.

***Country.*** Among the 84 studies, 57 reported the nationality of the participants and 27 did not. In all, 12 countries were represented across all the studies. The countries represented were France, Germany, Israel, Japan, Korea, Mexico, the Netherlands, New Zealand, United Kingdom, United States, Singapore, and Sweden. Several studies examined more than one population (e.g., Walters et al. 2011).

***Level of Analysis***. All 84 studies provided information regarding the level of analysis. All but one study focused on the individual level of analysis. The only exception was Salam et al. (2017), which focused on the group level of analysis.

## Section 2.7. Study Settings

Although all the studies were experimental lab studies, many were set up to mimic a particular real-world setting. Forty-four studies were designed to mimic a real-world setting and 40 studies were not. We identified five types of study setting. The first was robots in the home, where the setting mimicked a home environment. In these studies the robot helped or performed some type of domestic task other than care-giving. There were 18 studies in this category. The second was robot as caregiver. These studies were directed at understanding how the robot could be better designed to support health-related care. There were 15 studies in this category. The third setting was robots in an organizational work setting. In these studies the robot was placed in an office, manufacturing plant, or retail store setting. This category included 7 studies. The fourth setting was education. In this setting the robot supported some type of learning objective. There were only 3 studies in this category. Finally, the entertainment setting included robots attempting to entertain participants. There was only 1 study in this category. Please see figure 2.2.

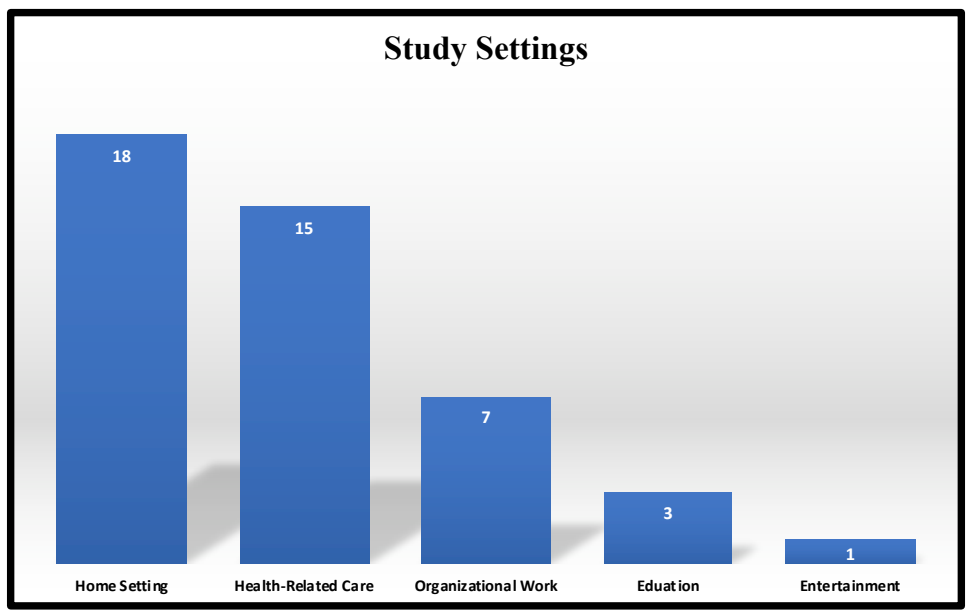

Figure 2.2 Publications by Study Setting

## Section 2.8. Type of Robot

The 84 studies together employed approximately 95 robots, several studies using two to three robots. At least 20 types of robots were used across all the studies. Many studies indicated whether the robot was a humanoid or a non-humanoid robot. Based on these studies' descriptions, 49 were humanoid robots and 15 were non-humanoid robots. Some studies also indicated what specific type of robot was used in the study. The Nao robot was by far the most used robot, identified across all studies. Eighteen studies employed a Nao robot. The second most used robot were the iCat and PeopleBot robots, with six studies each. The third most used robot was the Meka, with four studies. The fourth most used robot was the iCub, with three studies. The remaining 15 robots were used in one to two studies.

## Section 2.9. Robot Interaction Control

Robot interaction control is the method used to control the robot during its interaction with the human. Robot interaction is defined as the robot responses to the human over multiple human responses. This is in stark contrast to studies that either asked the human participant to watch the robot display a behavior independent of the human or watch the robot respond to someone or something other than the human participant. Five types of control were used to control the robots' interaction with the participant. The Wizard of Oz was the most frequently used robot control approach, with 24 studies indicating it as their approach. The Wizard of Oz approach pretends that the robot is autonomous but actually relies on a hidden human to control the robot remotely. The second most used robot interaction control approach was the automation autonomy, with 10 studies. Automation autonomy relies on the robot itself independently reacting to the human, based on the human's response. The third most used approach was the pre-programmed strategy, in six studies. The pre-programmed strategy pretends that the robot is autonomous but actually relies on a series of pre-defined robot responses that are not based on the human's response; that is, the robot's response is the same to every human regardless of the human's behavior. The third most used approach is a hybrid combination of the Wizard of Oz/automaton and autonomy, with four studies. This approach automates some of the robot behavior while allowing a human operator to control other behaviors. The remaining studies either did not indicate the type of robot interaction control or did not rely on any one type of interaction.

## Section 2.10. Thrust Area

We categorized the studies into four thrust areas based on You and Robert (2017). Thrust Area 1 was Human Personality and HRI and included studies focusing on the impact of human personality on outcomes of humans interacting with a robot. Thrust Area 2 was Robot Personality and HRI and included studies focusing on the impact of robot personality on outcomes of humans interacting with a robot. Thrust Area 3, Robot and Human Personality Similarities and Differences, focused on the impact of matching or mismatching human and robot personalities on outcomes of humans interacting with a robot. Finally, Thrust Area 4, Factors Impacting Robot Personality, included studies focusing on ways to invoke perceived robot personality.

## Section 2.11. Chapter Summary

In summary, chapter 2 presented and discussed the paper's literature search review. The selection criteria that was used to go from 220 articles to the 83 articles which comprised 84 studies was explained. In addition, chapter 2 also provided an overall summary of the publication venues, personality measures, outcomes, sample characteristics, study settings, type of robot and robot interaction control used across all the studies. Finally, chapter 2 concludes by listing the four thrust areas used to organize the literature.

# Chapter 3

# Thrust Area 1:

# Human Personality and Human–Robot Interaction

## Section 3.1. Inputs

Human personality has been an essential topic in the human–robot interaction literature on personality. Generally, most studies have assumed that human personality can be used to determine whether an individual would be more or less likely to interact with a robot and whether those interactions were likely to be enjoyable. Extraversion/introversion has been the most common human personality trait investigated in this thrust. We identified 34 articles in total that included human personality as an independent variable. Twenty-three of those articles examined the impact of extraversion/introversion traits.

Many studies employed extraversion/introversion because the literature on human-to-human interactions highlights its importance (Robert 2018). Many studies assume that humans who are more extraverted tend to be more social and should be more willing to interact with robots (Gockley and Mataric 2006; Haring, Watanabe, Silvera-Tawil, Velonaki, and Matsumoto 2015; Salem, Lakatos, Amirabdollahian, and Dautenhahn 2015; Syrdal, Dautenhahn, Woods, Walters, and Koay 2006; Syrdal, Koay, Walters, and Dautenhahn 2007). The literature also showed that individuals high in extraversion were more inclined to endow robots with a higher amount of trust than individuals with an introverted personality (Haring, Matsumoto, and Watanabe 2013). Furthermore, Ivaldi et al. (2017) examined whether the dynamics of gaze and speech were related to the extraversion. They found that extraverted individuals were more likely to talk to robots. Other studies employed human personality as proximity for the ability of humans to infer robot emotions from their body and facial expressions (Chevalier, Martin, Isableu, and Tapus 2015).

Although introversion/extraversion has been the most common personality trait in the literature of the human–robot interaction, researchers have also studied other personality traits. An example is openness to experience. Openness to experience was found to be an antecedent that promoted the acceptability of assistive robotic technologies (Conti, Commodari, and Buono 2017). Furthermore, the research of Takayama and Pantofaru (2009) indicated that personal experience with the robot reduced an individual's personal space around a robot. Moreover, Ogawa et al. (2009) showed that with rising openness of individuals, the agreeableness and extraversion ratings of the robot decreased.

On the other hand, a handful of research work focused on the conscientiousness trait. Cruz-Maya and Tapus (2016a) showed that conscientiousness is a predictor for task performance. They demonstrated that individuals high in conscientiousness performed better when they were reminded about their task by a robot rather than a low-conscientiousness individual. Regarding conscientiousness and the approach direction preferences, Syrdal, Koay et al. (2007) found that individuals low in conscientiousness were more inclined to allow the robot to approach closer than their counterparts who were high in conscientiousness.

Other studies focused on neuroticism. For instance, Damholdt et al. (2015) found neuroticism to be negatively correlated with  mental relatedness with the robot. That result was echoed in previous studies. Takayama and Pantofaru (2009) found that individuals high in neuroticism physically distanced themselves from the robot.

The measurement of human personality has been done in different ways. Most studies used Ten-Item Personality Inventory (TIPI) (e.g., Brandstetter, Beckner, Bartneck, and Benitez 2017; Salem et al. 2015; Sandoval, Brandstetter, Obaid, and Bartneck 2016; Sehili, Yang, Leynaert, and Devillers 2014). Other studies employed different scales like the Big Five Domain Scale form IPIP (Syrdal, Koay, et al. 2007), NEO Five Factor Inventory (Damholdt et al. 2015; Ogawa et al. 2009), Negative Attitudes toward Robots Scale (NARS; Nomura, Kanda, Suzuki, Kato 2008; Nomura, Shintani, Fujii, and Hokabe 2007) and the Eysenck personality questionnaire (Haring et al. 2013; Haring, Matsumoto, and Watanabe 2014; Haring et al. 2015). Table 3.1 presents a summary matrix table of human personality on human–robot interaction literature.

Table 3.1 Human Personality Inputs

| Article | Personality traits | Measure |
|---------|-------------------|---------|
| Bernotat and Eyssel 2017 | Yes/All Big Five (i.e. openness to experience, conscientiousness, extraversion, agreeableness, and neuroticism). | Big Five Inventory (BFI) |
| Brandstetter, Beckner, Bartneck, and Benitez 2017 | Yes/Agreeableness and openness to experience | Big Five personality via TIPI test |
| Cruz-Maya and Tapus 2016 a&b | Yes/Only neuroticism | Big 5 personality test (45 items) |
| Ogawa et al. 2009 | Yes/Only openness to experiences | NEO Five Factor Inventory questionnaire which only contains 60 items |
| Takayama and Pantofaru 2009 | Yes/All Big Five, need for cognition, NARS | Big Five by John and Srivastava 1999<br>Need for cognition by Cacioppo et al. 1984<br>NARS by Nomura et al. 2006 |
| Gockley and Matarić 2006 | Yes/Extraverted and introverted | 50-question Big Five Inventory  by Goldberg (1999) |
| Haring, Matsumoto, and Watanabe 2013 | Yes/Psychoticism, extraversion and neuroticism | Eysenck Personality Questionnaire (Japanese version) by Berg et al. 1995 |

| Walters et al. 2005 | Yes/Proactiveness, social reluctance, timidity and nervousness | Eysenck's Three-Factor Psychoticism, Extraversion and Neuroticism (PEN) model |
|---|---|---|
| Syrdal, Dautenhahn, Woods, Walters and Koay 2006 | Yes/Big Five | Big Five IPIP, 10 items |
| Salem, Lakatos, Amirabdollahian and Dautenhahn 2015 | Yes/Extraversion and emotional stability | Big Five personality via TIPI test Gosling et al. 2003 |
| Syrdal, Koay, Walters and Dautenhahn 2007 | Yes/Big Five | Big Five via IPIP (Goldberg 1999) |
| Chidambaram, Chiang, and Mutlu 2012 | Yes/Assertiveness | International Personality Item Pool Interpersonal Circumplex (IPIP-IPC) by Markey and Markey 2009 |
| Nomura, Kanda, Suzuki, and Kato 2008 | Yes/NARS & RAS | NARS & RAS (taken from the paper) |
| Ivaldi et al. 2017 | Yes/Extraversion and NARS | Big Five model via Goldberg 1990 NARS by Nomura et al. 2006 |
| Sandoval, Brandstetter, Obaid and Bartneck 2016 | Yes/Big Five | Big Five traits via TIPI by Gosling et al. 2003 |
| Park, Jin, and del Pobil 2012 | Yes/Extraverted vs. intermediate vs. introverted | Myers–Briggs Type Indicator (MBTI) |
| Nomura, Shintani, Fujii, and Hokabe 2007 | Yes/NARS & RAS | NARS & RAS (taken from the paper) |
| Haring, Matsumoto, and Watanabe 2014 | Yes/Big Five | Eysenck personality questionnaire (Japanese version) by Hosokawa et al. 1993 Godspeed questionnaire |
| Haring, Watanabe, Silvera-Tawil, Velonaki and Matsumoto 2015 | Yes/Extraversion | Eysenck personality questionnaire (Japanese version) by Hosokawa et al. 1993 Godspeed questionnaire |
| Cruz-Maya and Tapus 2016 a & b | Yes/Big Five | Big Five model via Goldberg 1990 |
| Sehili, Yang, Leynaert, and Devillers 2014 | Yes/Big Five | Big Five traits TIPI by Gosling et al. 2003 |
| Kleanthous et al. 2016 | Yes/Big Five | Big Five via IPIP Goldberg 1999 |
| Damholdt et al. 2015 | Yes/Big Five | NEO-EFI via Costa and McCrae 1992 |
| Kimoto et al. 2016 | Yes/Big Five | Big Five traits by John and Srivastava 1999 |
| Syrdal, Dautenhahn, Woods, Walters, | Yes/Big Five | Big Five via IPIP Goldberg 1999 |

| Koay 2007 | | |
|---|---|---|
| Walters, Syrdal, Dautenhahn, Te Boekhorst and Koay 2008 | Yes/Emotional stability, extraversion, agreeableness, conscientiousness and intellect | Big Five Model, International Personality Item Pool (IPIP) for humans; new survey for robots |
| Rosenthal-von der Pütten, Krämer, Hoffmann, Sobieraj and Eimler 2013 | Yes/Non-Big Five | UCLA Loneliness Scale; Saarbrueck Personality Questionnaire (SPF) and Mehrabian Affiliative Tendency |
| Weiss, Van Dijk and Evers 2012 | Yes/Extraversion and introversion | Wiggins personality test |
| Chevalier, Martin, Isableu and Tapus 2015 | Yes/Extraversion and introversion | Big Five personality traits via John and Srivastava 1999 |
| Vollmer, Rohlfing, Wrede and Cangelosi 2015 | Yes/Big Five | Big Five personality test via Rammstedt and John 2007 |
| MacDorman and Entezari 2015 | Yes/Neuroticism and anxiety and others | Neuroticism: 23BB5 (Duijsens & Diekstra 1995); Anxiety IPIP (Goldberg 1999); Personal Distress Interpersonal Reactivity Index (IRI by Davis 1980); Animal Reminder via Fear Survey Schedule-111 (Arrindell et al. 2003); Human_Robot Uniqueness Index (in paper); Religious Fundamentalism RRFS (Altemeyer and Hunsberger 2004); NARAS (Nomura et al. 2006) |
| Conti, Commodari, and Buono 2017 | Yes/Big Five | Big Five Questionnaire (BFQ-2), Italian adaptation of Caprara, Barbaranelli, Borgogni, and Secchione 2007 Eurobarometer Questionnaire by European Commission 2012 |
| Szczuka and Krämer 2016 | Yes/Loneliness, importance of social contacts, fear of rejection, the individual degree of interaction deficits, anthropomorphic tendency and the negative attitude toward | Anthropomorphism Questionnaire by Neave, Jackson, Saxton, and Hönekopp 2015; NARS scale by Syrdal, Dautenhahn, Koay and Walters 2009; UCLA Loneliness scale by Russel, Peplau and Cutrona 1980; Need to belong by Kramer et al. 2013; Social anxiety by Kolbeck 2008 |
| Reich-Stiebert and Eyssel (2015) | Yes/Need for cognition, | NARS by Nomura et al. 2006 |

| | NARS, anxiety | RAS by Nomura et al. 2006; NFC scale proposed by Bless et al. 1994 Technology Commitment scale by Neyer, Felber and Gebhardt 2012 |
|---|---|---|

## Section 3.2. Outcomes

Research in HRI has sought to understand the impact of human personality on the interaction with robots. What are the consequences of enacting certain personality traits? This is a vital question motivating scholars to undertake research in this area. In brief, researchers have found a positive impact of human personality on various outcomes in several aspects of human–robot interaction. To understand the impact of human personality traits on human–robot interaction, researchers have investigated various dependent variables. Table 3.2 shows studies on human personality traits categorized by outcomes. The following paragraphs discuss each of the areas.

First, distance and approaching direction are salient outcomes in the literature of human personality and human–robot interaction. For instance, Walters et al. (2005) showed that individuals who approached a robot or were being approached by a robot preferred approach distances comparable to those in normal social interactions between individuals. The study also demonstrated that proactive individuals were more likely to keep longer distance between themselves and robot. On the other hand, other studies found that negative attitudes and anxiety impacted the distance between the participants and the robot (Nomura et al. 2007). The results of Takayama and Pantofaru's (2009) work revealed that experiences with pets and robots decreased the personal space individuals maintain around robots. Moreover, the study showed that a robot's head that was oriented toward participants' face (versus their legs) impacted proxemic behavior. And women maintained larger spaces from robots than men did. In addition, the study demonstrated that individuals who were higher in agreeableness moved closer toward robots, whereas individuals who were higher in neuroticism stood farther from robots. Conversely, Syrdal et al. (2006) found that personality traits did not have any impact on approach distance. Furthermore, individuals who were higher in extraversion had a slightly higher tendency to tolerate robot behavior.

The second area pertains to the perceptions and attitudes toward the robot. For instance, Reich-Stiebert and Eyssel (2015) found that negative attitudes and robot anxiety decreased as a function of individuals' technology commitment. The study also found that demographic characteristics significantly impacted individuals' attitudes toward educational robots. Specifically, younger participants reported higher negative attitudes toward robots than older participants. Furthermore, women showed higher negative attitudes and less tendency to interact with educational robots than men did. Conversely, Chidambaram, Chiang, and Mutlu (2012) did not find support for the gender impacts on individuals' perceptions of the robot's persuasiveness. Additionally, Damholdt et al. (2015) examined an elderly population. The term *elderly* population in this study and across several others in this review (see Sehili et al. 2014 and Kleanthous et al. 2016) refers to a population that had 1) an average age of 80 and 2) were living in a retirement community. Damholdt et al. (2015) found that elderly people did not show any statistically significant change in their behavior toward robots whether they were informed about the ability of the robot to be tele-operated or not. The study reported that beliefs about robot autonomy and functionality did not impact elderly people's behavior toward the robot. Concerning human personality traits and perceptions toward robots, scholars found that extraversion, conscientiousness, and emotional stability were perceived approximately the same in robots and humans. Yet, individuals perceived robots as less open and agreeable than human agents (Sandoval et al. 2016). Furthermore, Bernotat and Eyssel (2017) found that openness to experience, extraversion, conscientiousness, and neuroticism did not predict a positive evaluation of the interaction with an intelligent robotics apartment. Yet, high agreeableness predicted the positive evaluation of the interaction with the robot.

Third, scholars studied the impacts of human personality on anthropomorphism, which is the attribution of human characteristics to an inhuman entity, in this case robots. Park, Jin, and del Pobil (2012) found that individuals' personality impacted their anthropomorphism toward the robot such that extraverts assigned a higher degree of human characteristics to the robot than introverts. The same results have been reported in other studies. For instance, Salem et al. (2015) found that individuals who were high in extraversion and emotional stability were more likely to anthropomorphize the robot more and feel closer to it. Damholdt et al. (2015) demonstrated that neuroticism and anthropomorphic thinking negatively correlated with mental relatedness.

The fourth area pertains to the trust in the human–robot interaction. Researchers explored the impact of personality traits on trust in the robot. Haring, Matsumoto, and Watanabe (2013 and 2014) found that extraversion was positively correlated with the amount of trust sent in a trust game. This was confirmed by a Salem et al. (2015) study, which found that extraverted individuals anthropomorphized the robot more and felt closer to it. Yet, extraversion and emotional stability did not impact individuals' trust development with regard to the robot.

The fifth set of constructs pertains to emotion toward the robot. Rosenthal-von der Pütten et al. (2013) found that individuals show emotional reactions toward robots. Yet, Chevalier et al. (2015) demonstrated that human personality traits do not impact the emotion recognition behavior.

Finally, scholars have investigated various topics such as robot acceptance. For instance, Conti, Commodari, and Buono (2017) examined the impact of human personality traits on robotics technology acceptance. The results emphasized the vital role of openness to experience and extraversion for promoting acceptability. Furthermore, a few studies examined the impact of robot embodiment on human–robot interaction. For instance, Cruz-Maya and Tapus (2016b) examined whether the presence of embodiment would promote human learning from the robot. In addition, Ogawa et al. (2009) investigated the influence of an agent's embodiment on the robot's persuasiveness. Last, learning and engagement with the robot were examined in the literature. Brandstetter, Beckner, Bartneck, and Benitez (2017) studied the impact of agreeableness and openness to experience on the likelihood of a person adopting the interlocutor's vocabulary choices. Cruz-Maya and Tapus (2016b) studied the impact of human traits on learning from the robot.

Table 3.2 Human Personality Outcomes

| Outcome | Paper | Measures |
|---------|-------|----------|
| Distance and approach direction | Haring, Matsumoto, and Watanabe 2013 | The position the subject put the chair during the interaction task. |
| | Walters et al. 2005 | Two measurements for the comfortable approach distance and two for the comfortable withdrawal distance. |
| | Syrdal, Koay, Walters and Dautenhahn 2007 | The distance ratings were achieved by human subjects approaching a stationary robot in a straight line. |
| | Nomura, Shintani, Fujii and Hokabe 2007 | Only robot–human approach distances Walters et al. 2005 |
| | Haring, Matsumoto and Watanabe 2014 | The position the subject put the chair during the interaction task |
| | Takayama and Pantofaru 2009 | The average and minimum distance the participant reached relative to the robot's base scanner |
| | Syrdal, Dautenhahn, Woods, Walters and Koay 2006 | This experiment was divided into four scenarios. In each of these scenarios the robot approached from a different direction |
| Trust | Haring, Matsumoto and Watanabe 2013 | The amount of money sent in a trust game |
| | Salem, Lakatos, Amirabdollahian and Dautenhahn 2015 | Trust was measured based on self-reported quantitative and qualitative questionnaire data as well as on behavioral data that assesses trust as the participants' willingness to cooperate with a robot. |
| | Haring, Matsumoto and Watanabe 2014 | The amount of money paid by participants. |
| Anthropomorphism | Haring, Matsumoto and Watanabe 2013 | Godspeed questionnaire Bartneck, Kulić, Croft, and Zoghbi 2009 |
| | Haring, Matsumoto and Watanabe 2014 | Godspeed questionnaire Bartneck, Kulić, Croft, and Zoghbi 2009 |
| | Damholdt et al. 2015 | A 10-item questionnaire was developed to assess anthropomorphic thinking |
| | Salem, Lakatos, Amirabdollahian and Dautenhahn 2015 | Godspeed questionnaire by Bartneck, Kulić, Croft, and Zoghbi 2009 |
| | Park, Jin, and del Pobil 2012 | Nine items adapted from Schifferstein and Zwartkruis-Pelgrim 2008 |
| Task performance | Gockley and Matarić 2006 | The amounts of time on each task |
| | Cruz-Maya and Tapus 2016a | Time on task |
| Persuasiveness | Ogawa et al. 2009 | By asking the participants before and after the persuasive speech how much they would be willing to pay for the product |
| | Chidambaram, Chiang, and Mutlu 2012 | A post-experiment questionnaire |
| Engagement | Gockley and Matarić 2006 | High engagement condition: the robot behaved as described, while in the low engagement condition the robot did not follow participants' arm movement at all |
| | Ivaldi et al. 2017 | Gaze and speech signals were used to evaluate the engagement. (1) The time spent by the participants looking at the robot for the whole duration of the interaction. (2) The sequence of gaze toward the robot |
| Touching the robot | Haring, Matsumoto, and Watanabe 2013 | The time was measured from the first request until the subjects made physical contact with the robot's hand |
| | Haring, Watanabe, Silvera-Tawil, Velonaki and Matsumoto 2015 | Dictionary of tactile gestures was adapted from Yohanan and MacLean 2012 |
| | Haring, Matsumoto, and Watanabe 2014 | Touch time was measured from the first request until the subjects made physical contact with the robot's hand |
| | Nomura, Kanda, Suzuki, and Kato 2008 | Elapsed time before subjects touched the robot's body after being encouraged to do so |
| Likeability | Haring, Matsumoto, and Watanabe 2014 | Godspeed questionnaire via Bartneck, Kulić, Croft, and Zoghbi 2009 |
| Compliance with | Chidambaram, Chiang, and Mutlu 2012 | The number of changes that the |

| the robot's suggestions | | participants made in their rankings. This measure increased with changes in the direction of the robot's suggestions and decreased with changes against the robot's suggestions |
|---|---|---|
| Perceptions and attitudes | Sandoval, Brandstetter, Obaid, and Bartneck 2016 | TIPI test adapted from Gosling, Rentfrow, and Swann Jr. 2003 |
| | Chidambaram, Chiang, and Mutlu 2012 | Their perceptions of the robot's social and intellectual characteristics |
| | Damholdt et al. 2015 | The Attitudes toward Social Robots Scale (ASOR-5) questionnaire |
| | Bernotat and Eyssel 2017 | Positive and negative affect scale obtained from Krohne, Egloff, Kohlmann, and Tausch 1996 |
| | Reich-Stiebert and Eyssel 2015 | Negative Attitudes toward Robots Scale; Robot Anxiety Scale; they developed 11 items that tapped participants' willingness to interact with robots in the future |
| Emotions | Chevalier, Martin, Isableu, and Tapus 2015 | The robot had facial features that enabled participants to express the four emotions present in the situations |
| | Rosenthal-von der Pütten, Krämer, Hoffmann, Sobieraj, and Eimler 2013 | Positive and Negative Affect Schedule (PANAS) |
| Safety | Haring, Matsumoto, and Watanabe 2014 | Godspeed questionnaire Bartneck, Kulić, Croft, and Zoghbi 2009 |
| Uncanny valley | MacDorman and Entezari 2015 | Was operationalized as higher ratings of eeriness and lower ratings of warmth |
| Appearance | Walters, Syrdal, Dautenhahn, Te Boekhorst, and Koay 2008 | Participants provided a set of ratings on a Likert scale for their preference for robots' appearance |
| Intention to buy | Szczuka and Krämer 2016 | Intention to buy a sex robot now or within the next 5 years |
| Learning from robot | Cruz-Maya and Tapus 2016a | Measured as completion time |
| Satisfaction | Sehili, Yang, Leynaert, and Devillers (2014) | Developed questionnaire. |
| Cooperation | Sandoval, Brandstetter, Obaid, and Bartneck 2016 | The number of cooperative behaviors in every set of prisoner's dilemma game |
| Robot Presence | Park, Jin, and del Pobil 2012 | Lee, Peng, Jin, and Yan 2006 |

| Acceptance | Conti, Commodari, and Buono 2017 | UTAUT questionnaire |

## Section 3.3. Study Method, Sample, Context and Robot Type

Conducting experiments has been the common methodology to study the impact of human personality on the human–robot interaction. All 34 studies we identified for the topic of human personality were conducted with lab experiments. Based on the data provided, the sample sizes were between 11 and 489 participants. Participants' ages ranged from 19 years to 85 years. The samples were 1120 males to 931 females. The participants involved in study were from Japan (7), France (4), U.S. (4), U.K. (4) , Netherlands (2), New Zealand (2), Korea (1), Portugal (1), Denmark (1), and Italy (1). The participants were university students, staff and older adults.

It was not always clear what type of robot we employed. Nonetheless, based on the available information the most common robots were: Nao (6), Peoplebot (4), Meka (2), icub (2) and Roomba (1). The most common way to control the robot was via Wizard of Oz (10), Autonomous (4), and predefined programmed scripts (4). The humanoid robot was identified within this thrust area as the most employed.

## Section 3.4. Findings

This section incorporates the variables discussed and presents the findings from the studies on the topic of human personality. The following sections present findings organized around the Big Five human traits.

Extraversion is the most examined human trait in the human–robot interaction studies. The results of the Gockley and Matarić's (2006) experiment work showed that extraverted individuals preferred the robot to stay closer. Syrdal et al. (2006) found that personality traits did not have any impact on approach distance. Yet, individuals who were higher in extraversion had a slightly higher tendency to tolerate robot behavior. These results were echoed in other studies. A study conducted by Syrdal, Koay et al. (2007) demonstrated that individuals who were higher in extraversion were more likely to allow the robot to approach closer. Furthermore, Salem et al. (2015) found that individuals who scored high in extraversion were more likely to anthropomorphize the robot more and feel closer to it. Haring, Matsumoto, and Watanabe (2013 and 2014) found that extraverted individuals were more likely to endow the android robot with a higher offer in the trust game. Furthermore, Ivaldi et al. (2017) found that individuals who were high in extraversion talked more and longer with the robot.

Regarding the comparison between extraversion and introversion, Cruz-Maya and Tapus (2016a) found that introverted individuals were more motivated by the robot to accomplish the task earlier than extraverted individuals. The study also found that individuals who were higher in conscientiousness were better being on time when they were reminded by a robot.

Various studies have been performed to understand how neurotic individuals interact with a robot. Takayama and Pantofaru (2009) performed an experiment to understand the relationship between neuroticism and proxemics. Their work showed that individuals who were higher in neuroticism stood farther from robots. On the other hand, Haring et al. (2015) found that higher levels of neuroticism were associated with a longer period of time of touching; yet, higher levels of extraversion were associated with a faster touch of the robot by individuals. In addition, Cruz-Maya and Tapus's (2016b) work showed that individuals' personality plays a crucial role in learning. They found that learning performance was higher for individuals who scored high in neuroticism than for those who scored low in neuroticism.

Prior studies shed little light on the impact of the agreeableness trait on human–robot interaction. Bernotat and Eyssel's (2017) work revealed that extraversion, openness to experiences, neuroticism, and conscientiousness did not predict a positive evaluation of the interaction with an intelligent robotics apartment. Yet, high agreeableness had a positive impact on interaction with the intelligent robotics apartment. This might be explained by trust. Authors found that individuals high in agreeableness were trusting. Prior studies showed that trust decreased perceived risk (Pavlou, 2003) and increased the intention of technology use and perceived usefulness. Furthermore, agreeableness was examined to understand proxemic behavior. Takayama and Pantofaru (2009) found that individuals who were high in agreeableness moved closer toward robots, whereas those high in neuroticism stood farther from robots.

A considerable body of research examined the impact of openness to experiences as a key human trait on human–robot interaction. Brandstetter et al. (2017) found that speakers who scored higher in openness to experiences were more likely to adopt different kinds of words introduced by their robot peer. The openness to experiences trait plays a fundamental role in how individuals perceive the personality of the robot. Ogawa et al. (2009) found that with the rising openness of the participants toward experiences, extraversion and agreeableness ratings for the persuasive agent (android) decreased.

The importance of gender across human personality studies on human interaction with the robot seems to vary. Syrdal, Koay et al.'s (2007) research demonstrated that individuals' gender had some impact on approach direction preferences. When the robot approached from the side, there was no difference between males and females. Yet, when the approaching was directly from the front, females allowed the robot to approach closer than males did. Nomura et al. (2008) found that the anxiety and negative attitudes toward robots impacted individuals' behavior toward robots (spending time with and touching the robots). The study indicated that male individuals with high negative attitudes and anxiety toward robots were more likely to avoid talking to and touching the robot. Conversely, Chidambaram, Chiang, and Mutlu (2012) did not find support for the gender impact on individuals' perceptions of the robot's persuasiveness.

Another trend is associated with the impact of age on human interaction with the robot. Sehili et al. (2014) found that the personality of elderly people influenced their interaction with robots. Yet, Damholdt et al. (2015) found that elderly people did not show any statistically significant change in their behavior toward robots whether they were informed about the ability of the robot to be tele-operated or they were not informed. In addition, the results showed that beliefs about robot autonomy and functionality did not impact elderly people's behavior toward the robot. See Table 3.3 for a synopsis of predictors and outcomes in previous studies on human personality and human–robot interaction.

Table 3.3 Human Personality: Predictors and Outcomes

| Predictors | Outcomes |
| --- | --- |
| **Human personality:** Bernotat and Eyssel 2017; Chidambaram, Chiang, and Mutlu 2012; Cruz-Maya and Tapus 2016 a & b; Damholdt, Nørskov, Yamazaki, Hakli, Hansen, Vestergaard, and Seibt 2015; Gockley and Matarić 2006; Haring, Matsumoto, and Watanabe 2013; Haring, Matsumoto, and Watanabe 2014; Haring, Watanabe, Silvera-Tawil, Velonaki, and Matsumoto 2015; Ivaldi, Lefort, Peters, Chetouani, Provasi, and Zibetti 2017; Nomura, Kanda, Suzuki, and Kato 2008; Ogawa, Bartneck, Sakamoto, Kanda, Ono, and Ishiguro 2009; Park, Jin, and del Pobil 2012; Takayama and Pantofaru 2009; Salem, Lakatos, Amirabdollahian, and Dautenhahn 2015; Sandoval, Brandstetter, Obaid, and Bartneck 2016; Sehili, Yang, Leynaert, and Devillers 2014; Syrdal, Dautenhahn, Woods, Walters, and Koay 2006; Syrdal, Koay, Walters, and Dautenhahn 2007; Walters, Dautenhahn, Te Boekhorst, Koay, Kaouri, Woods, and Werry 2005 | **Distance from the robot:** Haring, Matsumoto, and Watanabe 2013; Haring, Matsumoto, and Watanabe 2014; Nomura, Shintani, Fujii, and Hokabe 2007; Syrdal, Koay, Walters, and Dautenhahn 2007; Takayama and Pantofaru 2009; Walters et al. 2005 |
| | **Trust:** Haring, Matsumoto, and Watanabe 2013; Haring, Matsumoto, and Watanabe 2014; Salem, Lakatos, Amirabdollahian, and Dautenhahn 2015 |
| | **Anthropomorphism:** Damholdt, Nørskov, Yamazaki, Hakli, Hansen, Vestergaard, and Seibt 2015; Haring, Matsumoto, and Watanabe 2014; Park, Jin, and del Pobil 2012; Salem, Lakatos, Amirabdollahian, and Dautenhahn 2015 |
| **Gender:** Bernotat and Eyssel 2017; Cruz-Maya and Tapus 2016b; Syrdal, Koay, Walters, and Dautenhahn 2007 | **Learning from the Robot:** Brandstetter, Beckner, Bartneck, and Benitez 2017; Cruz-Maya and Tapus 2016b) |
| **Robot embodiment:** Cruz-Maya and Tapus 2016b; Ogawa, Bartneck, Sakamoto, Kanda, Ono, and Ishiguro 2009 | **Time on task:** Motivation to complete the task Cruz-Maya and Tapus 2016a; Gockley and Matarić 2006 |
| **Vocal cues, gestures, gazes:** Chidambaram, Chiang, and Mutlu 2012; Cruz-Maya and Tapus 2016b; Nomura, Kanda, Suzuki, and Kato 2008 | **Robot persuasiveness:** Chidambaram, Chiang, and Mutlu 2012; Ogawa, Bartneck, Sakamoto, Kanda, Ono, and Ishiguro 2009 |
| **Robot height:** Cruz-Maya and Tapus (2016a) | **Robot engagement with the task**: Gockley and Matarić 2006 |
| **Openness to experience:** Brandstetter, Beckner, Bartneck, and Benitez 2017 | **Touch behaviors:** Haring, Matsumoto, and Watanabe 2014; Haring, Watanabe, Silvera-Tawil, Velonaki, and Matsumoto 2015; Nomura, Kanda, Suzuki, and Kato 2008 |
| **Agreeableness:** Brandstetter, Beckner, Bartneck, and Benitez 2017 | **Likeability:** Haring, Matsumoto, and Watanabe 2014 |
| **Experience:** Haring, Matsumoto, and Watanabe 2013; Takayama and Pantofaru 2009 | **Engagement:** Ivaldi, Lefort, Peters, Chetouani, Provasi, and Zibetti 2017 |
| **Likeability:** Haring, Matsumoto, and Watanabe 2013 | **Compliance with the robot's suggestions**: Chidambaram, Chiang, and Mutlu 2012 |
| **NARS**: Ivaldi, Lefort, Peters, Chetouani, Provasi, and Zibetti 2017; Nomura, Shintani, Fujii, and Hokabe 2007 | **Preference for robot approach:** Syrdal, Dautenhahn, Woods, Walters, and Koay 2006 |
| **Touch behaviors:** Haring, Watanabe, Silvera-Tawil, Velonaki, and Matsumoto 2015 | **Robot personality:** Chidambaram, Chiang, and Mutlu 2012; Sandoval, Brandstetter, Obaid, and Bartneck 2016 |
| **Robot autonomy:** Syrdal, Koay, Walters, and Dautenhahn 2007 | |
| **Direction of approach:** Syrdal, Koay, Walters, and Dautenhahn 2007 | **Safety:** Haring, Matsumoto, and Watanabe 2014 |
| **Cultural norms:** Syrdal, Koay, Walters, and Dautenhahn 2007 | **Satisfaction:** Sehili, Yang, Leynaert, and Devillers 2014 |
| **Robot performance:** Salem, Lakatos, Amirabdollahian, and Dautenhahn 2015 | **Evaluation of the robot:** Bernotat and Eyssel 2017 |
| **Distance:** Cruz-Maya and Tapus 2016a | **Cooperations:** Sandoval, Brandstetter, Obaid, and Bartneck 2016 |
| **Robot personality:** Bernotat and Eyssel 2017; Park, Jin, and del Pobil 2012; Takayama and Pantofaru 2009 | **Psychological relatedness:** Damholdt, Nørskov, Yamazaki, Hakli, Hansen, Vestergaard, and Seibt 2015 |
| **Interaction with the robots:** Haring, Matsumoto, and Watanabe 2013; Haring, Matsumoto, and Watanabe 2014; Syrdal, Koay, Walters, and Dautenhahn 2007 | **Psychological closeness:** Salem, Lakatos, Amirabdollahian, and Dautenhahn 2015 |
| **Emotion:** Bernotat and Eyssel 2017; Nomura, Shintani, Fujii, and Hokabe 2007 | **Social presence of the robot**: Park, Jin, and del Pobil 2012 |
| **Agent(human/ robot):** Sandoval, Brandstetter, Obaid, and Bartneck 2016 | **Performance:** Cruz-Maya and Tapus 2016a |
| **Game strategy:** Sandoval, Brandstetter, Obaid, and Bartneck 2016 | **Acceptance**: Conti, Commodari, and Buono 2017 |
| **Types of robot:** Syrdal, Koay, Walters, and Dautenhahn 2007 | **Uncanny valley:** MacDorman and Entezari 2015 |
| **Commitment:** Bernotat and Eyssel 2017 | **Emotion**: Chevalier, Martin, Isableu, and Tapus 2015; Rosenthal-von der Pütten, Krämer, Hoffmann, Sobieraj, and Eimler 2013 |
| | **Appearance**: Walters, Syrdal, Dautenhahn, Te Boekhorst, and Koay 2008 |

## Section 3.5. Chapter Summary

In summary, thrust area 1: Human Personality and Human–Robot Interaction examines the impact of human personality on human robot interactions. This literature examines the impact of the human's personality on human robot interactions. Results showed that human personality does directly and indirectly influences the humans reactions to robots. In particular, extroverted humans tended to prefer interacting with robots more than introverted humans.

# Chapter 4

# Thrust Area 2:

# Robot Personality and Human–Robot Interaction

## Section 4.1. Inputs

Researchers have examined robot personality based on human personality characteristics. Notably, extraversion is the most used personality trait in HRI research, mainly in two ways. First, extraversion has been studied in comparison to introversion. Many of the studies we reviewed simulated two types of robots, each of which was either extraverted or introverted, and asked people to rate which personality the robots had. For example, Lohse et al. (2008) studied whether people perceived distinctive characteristics of extraverted and introverted robots from each other. Likewise, Walters et al. (2011) investigated whether people recognized differences between robots displaying either extravert or introvert characteristics.

Also, studies have used extraversion as one of the Big Five personality traits, along with agreeableness, openness, conscientiousness, and neuroticism, to examine robot personality. For example, Hwang, Park and Hwang (2013) utilized the Big Five dimensions to examine affective interaction between humans and robots. Chee, Taezoon, Xu, Ng, and Tan (2012) also used the five traits to understand extraversion and agreeableness (e.g., friendliness and coldness) of robots. The researchers of these studies showed different physical appearances of robots (e.g., varied combinations of heads, trunks, and limbs, and presentation types [visual vs. physical], Hwang, Park, and Hwang, 2013; design features of robot images, Chee et al., 2012) and examined the relationships with the personality of robots. Besides, Hendriks, Meerbeek, Beoss, Pauws, and Sonneveld (2011) used the five dimensions to measure distinctive personality characterstics of animations (e.g., movements, sounds) for developing vacuum cleaner robots.

The measurement scales that have been used in HRI research are varied. The studies compared extraversion and introversion of robot personality employing Wiggins' (1979) scale of International Personality Item Pool (IPIP), for example Tay, Jung, and Park (2014); Leuwerink (2012); Weiss, van Dijk, and Evers (2012); and Windhouwer (2012). Adjectives that participants named to describe different kinds of robots (e.g., active–passive, interested–indifferent, talkative–quiet) were used to distinguish extraverted and introverted robots (e.g., Lohse et al., 2008). In the studies that employed the Big Five personality measures, IPIP (e.g., Chee et al., 2012) and TIPI Test (Gosling et al., 2003; e.g., Hendriks et al., 2011; Sandoval et al., 2016) were used.

In addition to extraverted personality, social ability of robots has been studied as part of robot personality. For example, to examine social intelligence of robots, de Ruyter, Saini, Markopoulos, and Van Breemen (2005) developed the social behaviors questionnaire (SBQ) by adapting items that reflected affective and social responses to others from the IPIP questionnaire (Goldberg, 1992; 1999) and implemented aspects of social intelligence into home dialogue robots. Looije, Neerincx and Cnossen (2010) also found social behaviors and characters played an important role of senior-assistant robots by implementing an ability to communicate high-level dialogue, use natural cues such as eye-gazing, and express appropriate emotions. Table 4.1 illustrates personality trait inputs from the literature.

Table 4.1 Robot Personality Inputs

| Article | Robot Personality Traits | Measures |
|---|---|---|
| Ogawa et al. 2009 | Extraversion, openness, agreeableness | NEO Five Factor Inventory (Japanese Property-Based Adjective Measurement questionnaire) |
| Moshkina and Arkin 2005 | Five factor: extraversion, agreeableness, conscientiousness, emotional stability (neuroticism) and openness to experience/intellect | Brief version of Goldberg's Unipolar Big-Five Markers (personality questionnaire) Saucier, 1994 |
| Looije et al. 2010 | Extraversion, openness to experience, emotional stability, agreeableness and conscientiousness | Big-Five: Goldberg 1992; Van Vliet 2001 |
| Tay, Jung, and Park 2014 | Perceived extraversion (as a manipulation check) | 10 items including cheerful, extraverted, and enthusiastic Wiggins 1979 |
| Windhouwer 2012 | Extraverted/introverted | Wiggins 1979 |
| Leuwerink 2012 | Extravert/introvert | Wiggins 1979 Lee et al. 2006 |
| Walters et al. 2011 | Extraverted/introverted: active, passive, interested, indifferent; talkative, quiet | Self-developed: The scale consisted of 14 adjective pairs from their pre-test that participants used to describe robot behavior |
| Lohse et al. 2008 | Extravert/introvert | Self-developed: based on adjectives pre-test participants named |
| Gu, Kim and Kwon 2015 | Extraverted or introverted | Not listed |
| Park et al. 2012 | Extraverted vs. intermediate vs. introverted | Not listed; Paper made the different levels of extraversion by facial movements and expressions of robots |
| Sundar et al. 2017 | Robot demeanor (seriousness/playfulness) | Self-developed: "Would you say that the robot's demeanor was more serious or more playful?" |
| Johal, Pesty and Calvary 2014 | Permissiveness and authoritativeness: dominance (demanding behavior, discipline, and punishment) and responsiveness (love, warmth, attention) in parents' parenting styles | Pleasantness, arousal, dominance (PAD) scale Russell and Mehrabian 1977 SAM representation Bradley and Lang 1994 |
| Kaniarasu and Steinfeld 2014 | Blame personality: User-/self-/team-blame | Groom et al. 2010 |
| Powers and Kiesler 2006 | Sociability, knowledge, dominance, human-likeness, masculinity, machine-likeness | Bem 1976 |

## Section 4.2. Outcomes

What are the impacts of robot personality on human–robot interaction? Researchers have been interested in the influences of different robot personality traits and the interests fall into mainly three categories: (1) the perception of robot (e.g., usefulness, trust, robot intelligence/capability); (2) intention to use robots with personality (intention to use/acceptance); and (3) the quality of interaction with robots (enjoyment, fun, perceived control).

First, research has indicated that different personality characteristics of robots influence trustworthiness, intelligence, capability, and perceived persuasiveness of robots. For instance, Windhouwer (2012) measured intelligence of extraverted and introverted robots in several task contexts. Looije et al. (2010) tested trustworthiness and empathy of robots depending on whether robots had social characteristics that were implemented with emotional behaviors and natural communication cues in the physical and virtual interaction settings. Ogawa et al. (2009) assessed perceived persuasiveness of robotic shopping agents based on likeness of the robots' physical appearances and personality to human originals.

In addition, the acceptance of robots has been the major outcome of different robot personality traits. For example, Tay, Jung, and Park (2014) examined acceptance of introverted and extraverted health care robots along with robot gender differences. De Ruyter et al. (2005) also measured satisfaction and acceptance of robots based on whether the robots were socially intelligent. Overall, research suggests that people have a high level of acceptance of robots that are considered to be extraverted and socially intelligent.

Additionally, one of the major topics in HRI has been the quality of interaction with robots that have different types of personality. What robot personality provides enjoyment and fun interacting with the robots? Park, Jin, and del Pobil (2012) examined the impacts of extravert and introvert robots on the perception of their friendliness, immersive tendency, and social presence. Meerbeek, Hoonhout, Bingley, and Terken (2006, 2008) investigated the influence of robot personality on enjoyment and perceived control of interaction with robots. Table 4.2 summarizes the robot personality variables measured in the literature, and Table 4.3 summarizes research on the dependent variables.

Table 4.2 Robot Personality Outcomes

| Personality | Paper | Robot Personality Dimensions | Measures |
|---|---|---|---|
| Big Five | Meerbeek et al. 2006 | Extraversion, agreeableness, and conscientiousness, neuroticism, openness to experience | Big Five: Boeree 2004 |
| | Meerbeek et al. 2008 | Extraversion, agreeableness, and conscientiousness, neuroticism, openness to experience | Big Five: Boeree 2004 |
| | Ogawa et al. 2009 | Extraversion, openness, agreeableness | NEO Five Factor Inventory (Japanese Property-based Adjective Measurement questionnaire) |

| | Looije et al. 2010 | Extraversion, openness to experience, emotional stability, agreeableness and conscientiousness, acceptance, empathy, conversational behavior | Big-Five: Goldberg 1992, empathy, trust: de Ruyter et al., 2005; acceptance: Venkatesh et al., 2003 |
|---|---|---|---|
| | Sandoval et al. 2016 | Extraversion, agreeableness, conscientiousness, neuroticism or emotional stability and openness | Big Five: TIPI Test Gosling et al. 2003 |
| | Tay, Jung and Park 2014 | Affective evaluations: Love/hateful, delighted/sad, happy/annoyed, calm/tense, excited/bored, relaxed/ angry, acceptance/disgusted, joy/sorrow | Crites et al. 1994 |
| | Hwang, Park and Hwang 2013 | Extraverted (sociable, outgoing, confident); agreeable (friendly, nice, pleasant); conscientiousness (helpful, hard-working); antineurotic (emotionally stable, adjusted); open (intelligent, imaginative, flexible) | Big-Five (no citation) |
| | Hendriks et al. 2011 | Neuroticism, extraversion, openness to experience, agreeableness and conscientiousness (calm, cooperative, efficient, likes routines, polite, systematic) | Big Five/NEO PI-R: Costa and McCrae 1992 TIPI: Gosling et al. 2003 |
| | Chee et al. 2012 | Perceived extraversion, agreeableness, conscientiousness, neuroticism, and openness | International Personality Item Pool (IPIP) |
| Wiggins Personality Test | Groom et al. 2009 | 1. Robot friendliness, robot integrity, robot malice 2. Extraverted, outgoing | 1. Modified version of Wiggins personality test (used to calculate an trait o index) 2. Self-developed as part of robot friendliness |
| | Windhouwer 2012 | Extravert and introvert (silent, shy, introverted, bashful, inward, unrevealing, unsparkling, undemonstrative, outgoing, extraverted, vivacious, jovial, enthusiastic, cheerful, perky and un-shy) | Wiggins personality test |
| | Weiss, van Dijk, and Evers 2012 | Extraverted/introverted | Wiggins personality test |
| | Leuwerink 2012 | Extraversion, introversion (with enjoyability, intelligence, fun) | Mixed Wiggins scale with Lee et al. 2006 |
| Non-Big Five | Sundar et al. 2017 | 1. Robot's eeriness 2. Robot's anxiety | 1. Eeriness: Ho and MacDorman 2010: "reassuring/eerie," "natural/freaky," and "ordinary/supernatur al" 2. Anxiety: Bartneck et |
| | | | al. 2009 |
| | Yamashita et al. 2016 | 23 pairs of adjectives including active or passive | Personality impression questionnaire (PIQ) Mori et al. 2012, Inoue and Kobayashi 1985, Endo et al. 2010 |
| | Broadbent et al. 2013 | 19 pairs of traits including warm-cold (three factors -- sociable, amiable, trustworthy -- extracted) | Asch's checklist of characteristics (1946) |
| | Lohse et al. 2008 | Active, passive; interested, indifferent; talkative, quiet; intelligent, stupid; predictable, unpredictable; consistent, inconsistent, fast, slow; polite, impolite; friendly, unfriendly; obedient, disobedient; diversified, boring; attentive, inattentive | Self-developed: based on adjectives pre-test participants named |

Table 4.3 Outcomes other Robot Personality

| Dependent Variables | Papers | Measures |
|---|---|---|
| Preferences | Park et al. 2012 | Nass and Lee 2001 |
| | Looije et al. 2010; preference for dialogues with robots | Self-developed: Which personal assistant's dialogue did they prefer? |
| Acceptance | Ogawa et al. 2009 (persuasiveness of presentation of the robot) | Self-developed persuasion questionnaire (on the argument, conservativeness, interestingness and price) |
| | Tay, Jung, and Park 2014 | Heerink et al. 2010 |
| | De Ruyter et al. 2005 | Venkatesh et al. 2003 |
| | Looije et al. 2010 | Venkatesh et al. 2003 |
| | Sundar et al. 2017 | Intention to further use: Venkatesh 2000: "I would like to have further interactions with the robot," "I would like to own a robot like the one I interacted with," and "I would recommend my friends to get a robot like the one I interacted with if it were available on the market" |
| Willingness to spend time with robots | Weiss, van Dijk and Evers 2012 | No citation |
| Accepting robot's advice | Powers and Kiesler 2006 | No citation |
| Compliance with robot | Weiss, van Dijk and Evers 2012 | No citation |
| Perceived control | Meerbeek et al. 2006 | Hinds 1998 |
| | Meerbeek et al. 2008 | Hinds 1998 |
| | Tay, Jung, and Park 2014 | Venkatesh 2000: Perceived behavioral control |
| Social presence | Park et al. 2012 | Lee et al. 2006 |
| Anthropomorphism | Park et al. 2012 | Schifferstein and Zwartkruis-Pelgrim 2008 |
| | Chee et al. 2012 | Bartneck et al. 2008: "fake/natural," "machinelike/humanlike," "unconsciousness/consciousness," "artificial/lifelike," and "moving rigidly/moving elegantly" |

| | Hendriks et al. 2011 | Think aloud while watching a video of robots; if participants called the robots he/him |
|---|---|---|
| Immersive tendency | Park et al. 2012 | Singer 1998 |
| Human-likeness | Yamashita et al. 2016 | As one of the adjective pairs (personality impression questionnaire: PIQ; Endo et al. 2010, Inoue and Kobayashi 1985, Mori et al. 2012) humanlike/machinelike |
| | Broadbent et al. 2013 | Self-developed: "How humanlike did you think this robot was?'' (very machine-like "0" to very humanlike "100") |
| Trust | Tay, Jung, and Park 2014 | Heerink et al. 2010 |
| | Groom et al. 2009 | Self-developed: (as part of *Friendliness* items) "helpful," "honest," "pretenseless," "reliable," and "trustworthy" |
| | Looije et al. 2010 | SBQ (de Ruyter et al. 2005) |
| | Kaniarasu and Steinfeld 2014 | Muir (1983) |
| | Weiss, van Dijk, and Evers 2012 | No citation |
| Recommendation appreciation | Meerbeek et al. 2006 | Self-developed. Accuracy and coverage: (1) "I appreciate the recommendations of Catherine/Lizzy" (direct); (2) "I would like to watch the programs that Catherine/Lizzy recommends" (accuracy); (3) "I think Catherine/Lizzy recommends all programs that could be of interest to me" (coverage) |
| | Meerbeek et al. 2008 | Self-developed. Accuracy and coverage: (1) "I appreciate the recommendations of Catherine/Lizzy" (direct); (2) "I would like to watch the programs that Catherine/Lizzy recommends" (accuracy); (3) "I think Catherine/Lizzy recommends all programs that could be of interest to me" (coverage) |
| Attachment | Moshkina and Arkin 2005 | Self-developed: "With every session, I was getting more attached to the dog" |
| Empathy | Looije et al. 2010 | SBQ (de Ruyter et al. 2005) |
| Satisfaction | de Ruyter et al. 2005 | User satisfaction with consumer products (de Ruyter and Hollemans 1997) |
| | Windhouwer 2012 | As part of fun (Isbister and Nass 2000; enjoyable, exciting, fun and satisfying) |
| | Lohse et al. 2008 | Self-developed: "How satisfied are you with the robot's behavior?" |
| Enjoyment | Meerbeek et al. 2006 | Huang et al. 2001 |
| | Meerbeek et al. 2008 | Huang et al. 2001 |

| | Moshkina and Arkin 2005 | Self-developed: "I enjoyed the interaction with the robotic dog" |
|---|---|---|
| | Gu, Kim, and Kwon 2015 | Self-developed: "This exhibition through the robot docent is very enjoyable" |
| | Leuwerink 2012 | Mixed Wiggins scale with Lee et al. 2006 |
| | Windhouwer 2012 | Lee et al. 2006: enjoyable, fun and entertaining |
| | Weiss, van Dijk and Evers 2012 | No citation |
| Fun | Leuwerink 2012 | Mixed Wiggins scale with Lee et al. 2006 |
| | Windhouwer 2012 | Isbister and Nass 2000 (enjoyable, exciting, fun and satisfying) |
| | Weiss, van Dijk and Evers 2012 | No citation |
| Likeability | Yamashita et al. 2016 | As one of the adjective pairs (personality impression questionnaire: PIQ; Endo et al. 2010, Inoue and Kobayashi 1985, Mori et al. 2012): desirable/undesirable; adorable/weird |
| | Lohse et al. 2008 | Self-developed: "How much do you like the robot?" |
| | Looije et al. 2010 | Self-developed: How much they liked each personal assistant |
| | Kaniarasu and Steinfeld 2014 | Groom et al., 2010 |
| Friendliness | Groom et al. 2009 | Self-developed: "cheerful," "enthusiastic," "extraverted," "happy," "helpful," "kind," "likeable," "outgoing," and "warm" |
| | Park et al. 2012 | Groom et al. 2010; Park et al. 2011 |
| | Hwang, Park and Hwang 2013 | As part of Big Five personality |
| | Yamashita et al. 2016 | As one of the adjective pairs (personality impression questionnaire: PIQ; Endo et al. 2010, Inoue and Kobayashi 1985, Mori et al. 2012): friendly/unfriendly |
| | Lohse et al. 2008 | As part of personality (self-developed: based on adjectives pre-test participants named) |
| Animacy | Chee et al. 2012 | Bartneck et al. 2008: "dead/alive," "stagnant/lively," "mechanical/organic," "artificial/lifelike," "inert/interactive," and "apathetic/responsive" |
| | Hendriks et al. 2011 | Think aloud while watching a video of robots; if participants said it was alive or like a domestic animal, a dog or an infant |
| | Broadbent et al. 2013 | Self-developed: "Did the robot seem alive?" (not at all alive "0" to very much alive "100") |
| Cooperativeness | Sandoval et al. 2016 | Set of Prisoner's Dilemma and the Offer made in Ultimatum Game |

| | Hendriks et al. 2011 | As part of Big Five: NEO PI-R Costa and McCrae 1992 TIPI Gosling et al. 2003 |
|---|---|---|
| Reciprocations | Sandoval et al. 2016 | Set of Prisoner's Dilemma and the Offer made in Ultimatum Game |
| Warmth | Groom et al. 2009 | Self-developed: (as part of *Friendliness* items) "cheerful," "enthusiastic," "extraverted," "happy," "helpful," "kind," "likeable," "outgoing," and "warm" |
| | Looije et al. 2010 | As part of social personality of robot (warm, creative, talkative, original, spontaneous, artistic) |
| Robot social intelligence | De Ruyter et al. 2005 | Items selected from the social behaviors questionnaire (SBQ): altruism, assertiveness, competence dutifulness, empathy, helpfulness modesty, responsibility, sociability sympathy, trust |
| Robot intelligence | Tay, Jung, and Park 2014 | Crites et al. 1994: wise/foolish |
| | Sundar et al. 2017 | Bartneck et al. 2009: "incompetent/competent," "ignorant/knowledgeable," "irresponsible/ responsible," "unintelligent/intelligent," and "foolish/sensible" |
| | Hwang, Park, and Hwang 2013 | As part of openness (Big Five personality: intelligent, imaginative, flexible) |
| | Leuwerink 2012 | Mixed Wiggins scale with Lee et al. 2006 |
| | Windhouwer 2012 | Self-developed: intelligent and clever |
| | Weiss, van Dijk, and Evers 2012 | No citation |
| | Lohse et al. 2008 | As part of personality (self-developed: based on adjectives pre-test participants named) |
| Robot capability | Tay, Jung, and Park 2014 | Crites et al. 1994 capable/incapable |
| | Johal, Pesty, and Calvary 2014 | Effectiveness via Bartneck et al 2008 |
| Ease of use/usefulness | Meerbeek et al. 2006 | Ease of use via Venkatesh and Davis 2000 usefulness via Van der Heijden 2004 |
| | Meerbeek et al. 2008 | Ease of use via Venkatesh and Davies 2000 usefulness via Van der Heijden 2004 |
| | Walters et al. 2011 | Self-developed: The scale consisted of 14 adjective pairs from their pre-test that participants used to describe robot behavior |
| | Moshkina and Arkin 2005 | Self-developed: It was easy to get the robotic dog perform the commands; It was easy to understand whether the robotic dog was performing the command or not |
| | Lohse et al. 2008 | As part of personality (self- |

| | | developed: based on adjectives pre-test participants named) |
|---|---|---|
| | Tay, Jung, and Park 2014 | Crites et al. 1994: useful/useless |
| | Hwang, Park, and Hwang 2013 | As part of favorable emotion toward robots (useful, relaxing, safe, accessible, amiable; Kanda et al. 2001, Mitsunaga et al. 2008, Scopelliti et al. 2005 |
| Robot safety | Chee et al. 2012 | Anxious/Relaxed<br>Agitated/Calm<br>Quiescent/Surprised<br>Bartneck et al. 2008 |
| | Tay, Jung, and Park 2014 | Crites et al. 1994: safe/unsafe |
| | Hwang, Park, and Hwang 2013 | As part of favorable emotion toward robots (useful, relaxing, safe, accessible)<br>Kanda et al. 2001<br>Mitsunaga et al. 2008<br>Scopelliti et al. 2005 |
| | Yamashita et al. 2016 | As one of the adjective pairs (personality impression questionnaire [PIQ];<br>Endo et al. 2010,<br>Inoue and Kobayashi 1985<br><br>Safe/Dangerous<br>Mori et al. 2012 |
| Robot's eeriness | Sundar et al. 2017 | Reassuring/Eerie<br>Natural/Freaky<br>Ordinary/Supernatural<br>Ho and MacDorman 2010 |
| | Broadbent et al. 2013 | Self-developed: "How eerie did the robot's face look?" (not at all eerie "0" to very eerie "100" for the human and silver faces only) |
| Social attraction | Sundar et al. 2017 | I think this robot could be a friend of mine<br>I think I could have a good time with this robot<br>I would like to spend more time with this robot<br>Lee et al. 2006 |
| Sense of team | Groom et al. 2009 | Self-developed: "I felt that the robot and I were a team" |

## Section 4.3. Study Method, Sample, Context

To understand components of robot personality and its effect, all of the 25studies we reviewed took a design and experimental approach. Researchers first tended to conduct a literature review to derive components of human personality and apply them to robot personality. For example, based on the Big-Five-based scales such as NEO PI-R (Costa and McCrae 1992) researchers surveyed what constructs robot personality (e.g., Hendriks et al. 2011). Then, they developed prototypes of robots reflecting those traits and experimented to find whether the robots were perceived as having the intended traits, and the effects of the perceived personality. For prototyping, videos containing an animated robot with certain personality characteristic were often used (e.g., Broadbent et al. 2013, Lohse et al. 2008, Walters et al. 2011).

Research on robot personality was conducted mostly in the home setting (8), healthcare setting (5) followed by the organizational work settings (2). One of the most distinctive contexts was health care services at home. For example, robot personality and its impacts were measured when robots carried out daily health management for diabetes prevention (Looije et al. 2010) and blood pressure checking (Broadbent et al. 2013). Based on the data provided, the sample sizes were between 12 and 200 participants. The samples were 511 males to 504 females. The participants involved in study were from Germany (5), Netherlands (3), U. K. (2), Singapore (2), Korea (2), Mexico (2), Japan (2), Sweden (1), U. S. (1), and New Zealand (1). The type of robot employed was not always clear. However, based on the available information Nao (4) and icat (4) were used more often followed by Peoplebot (1). The most common way to control the robot was via Wizard of Oz (8), Autonomous (3), and predefined programmed scripts (3).

## Section 4.4. Findings

In summary, research on robot personality has sought to investigate what forms a distinct robot personality, whether people perceive the personality of robots as designed, and what impacts robot personality have on attitudes toward robots. Research has used human personality measures such as the Big Five scale to understand people's perception of robot personality, and extraversion and introversion of robots were the characteristics most often examined. Research has also found which personalities of robots were more often preferred in terms of acceptableness, trustworthiness, enjoyableness, and ease of use . Table 4.4 has a synopsis of predictors and outcomes from previous studies on robot personality.

However, some of the studies that we reviewed showed contrasting findings regarding the perception of robot personality in different contexts of human–robot interaction. First, the different types of tasks and roles for which robots were used affected perceived robot personality. Compared to many studies that examined robot personality in a single context (e.g., mostly for companion and health assistance), Windhouwer et al. (2012) examined introverted and extraverted robots for many different roles (e.g., chief executive officer, pharmacist, and teacher) and showed the different effects robot personality had on perceived intelligence, fun, and enjoyableness of the robots. Sundar et al. (2017) found that a robot's assistant job and companion job required different aspects of robot personality. Tay, Jung, and Park (2014) also implemented robots in two occupations (security vs. health care) and tested what personality traits were more preferred for each of the jobs.

Additionally, Weiss, van Dijk, and Evers (2012), in their preliminary study, assumed that the cultural background of users might influence personality attributions of robots and expectation and preferences of robots.

Table 4.4 Robot Personality: Predictors and Outcomes

| Predictors | Outcomes |
|---|---|
| **Extraversion vs. introversion:** Gu, Kim, and Kwon 2015; Leuwerink 2012; Lohse et al. 2008; Park, Jin, and del Pobil 2012; Tay, Jung, and Park 2014; Walters et al. 2011; Weiss, van Dijk, and Evers 2012; Windhouwer 2012 | **Robot personality:** Chee et al. 2012, Sandoval et al. 2016 |
| | **Trust:** Groom et al. 2009; Looije et al. 2010; Sandoval et al. 2016; Tay, Jung, and Park 2014 |
| **Big Five personality traits**: Hendriks et al. 2011; Hwang, Park, and Hwang 2013; Meerbeek et al. 2006, 2008; Moshkina and Arkin 2005; Ogawa et al. 2009 | **Robot intelligence/capability:** Hwang et al. 2013; Leuwerink 2012; Lohse et al. 2008; Sundar et al. 2017; Tay et al. 2014; Weiss et al. 2012; Windhouwer 2012 |
| **Social intelligence**: de Ruyter et al. 2005; Looije, Neerincx, and Cnossen 2010 | **Perceived persuasiveness/recommendation appreciation:** Looije et al 2010; Ogawa, Bartneck and Sakamoto 2009; Powers and Kiesler 2006 |
| Friendliness, integrity, malice: Groom et al. 2009 | **Acceptance/Intention to use robots**: de Ruyter et al. 2005; Looije et al. 2010; Meerbeek et al. 2008; Sundar et al. 2017; Tay et al. 2014 |
| **Authoritativeness**: Johal, Pesty, and Calvary 2014 | |
| | **Perceived control:** Meerbeek et al. 2006, 2008 |

## Section 4.5. Chapter Summary

In summary, thrust area 2: Robot Personality and Human–Robot Interaction examines the impact of robot personality on human robot interactions. This literature examines the impact of the robot's personality on human robot interactions. Results showed that the type of personality displayed by the robot directly and indirectly influences the degree of fun and enjoyableness humans have with robots. In particular, humans tended to prefer robots that displayed a extroverted personality. This may be due the social nature of most of the tasks used in the studies. Nonetheless, as a collection the studies clearly highlight the importance of the robot's personality.

# Chapter 5

# Thrust Area 3:

# Robot and Human Personality Similarities and Differences

## Section 5.1. Inputs

Similarities and differences between a robot and a human have been an essential topic in the human–robot interaction literature. Most research in this topic has taken a binary approach to capturing the similarity and differences of personalities—whether or not the personalities between a robot and a human are similar. In particular, many studies measure an individual's personality and have him or her interact with a robot that demonstrates either a similar or the opposite personality trait. For instance, Celiktutan and Gunes (2015) measured participants' introversion/extraversion based on Big Five personality traits (Paunonen and Ashton 2001) and presented a robot with one of the two personalities. Similarly, Niculescu and colleagues (2013) employed a way to implement a personality match between a robot and a human. In their study, individuals were to compare their own personality in terms of extraversion with a robot's perceived extraversion. They varied the robot's voice pitch to express extraversion, such that the extraverted robots had higher-pitch voices, while the introvert robots had lower-pitch voices. See Table 5.1 for human personality inputs.

Table 5.1 Human Personality Inputs

| Article | Personality Traits | Measure |
|---|---|---|
| Celiktutan and Gunes 2015 | Extraversion and introversion | All Big Five traits/Big Five Inventory (BFI) |
| Cruz-Maya and Tapus 2017 | Extraversion and introversion | Big Five, Extraversion: Goldberg, 1990 |
| So et al. 2008 | Extraversion and introversion Thinking and feeling | Myers–Briggs Type Indicator (MBTI) Murray 1990 |
| Lee et al. 2006 | Extraversion and introversion | Myers–Briggs Type Indicator (MBTI) Murray 1990 |
| Tapus and Matarić 2008 | Extraversion and introversion | Eysenck Personality Inventory (EPI) |
| Joosse et al. 2013 | Extraversion and introversion | International Personality Item Pool (IPIP) Wiggins 1979 |
| Aly and Tapus 2013 | Extraversion and introversion | Not specified |
| Groom et al. 2009 | Not for independent variables. Participant personalities were measured for a control variable | International Personality Item Pool (IPIP) Wiggins 1979 |
| Salam et al. 2017 | Extraversion and introversion | All Big Five traits/Big Five Inventory (BFI) |
| Andrist et al. 2015 | Extraversion vs. introversion, and motivation | Big Five by John and Srivastava 1999 |

| | | Intrinsic and extrinsic motivation by Guay et al. 2003 |
|---|---|---|
| De Graaf and Ben Allouch 2014 | All Big Five personality traits | All Big Five traits/Big Five Inventory (BFI) |
| Mileounis et al. 2015 | Dominance and extraversion | International Personality Item Pool (IPIP) Wiggins 1979 |
| Niculescu et al. 2013 | Extraversion and introversion | Myers–Briggs Type Indicator (MBTI) Murray 1990 |
| Dang and Tapus 2015 | Extraversion and introversion | All Big Five traits/Big Five Inventory (BFI) |
| Aly and Tapus 2016 | Extraversion and introversion | Not specified |

Indeed, extraversion/introversion has been the most common dimension used to examine the personality match between a robot and a human. We identified fifteen papers in total concerning the human–robot personality match and found that twelve studies measured extraversion/introversion as a scale to capture the human–robot personality match. In addition to the studies mentioned above, Andrist, Mutlu, and Tapus (2015) reported that a robot's extraversion was manipulated by changing the eye movements of robots. Specifically, the longer and the more frequent gaze behavior toward a human counterpart by a robot led the human to perceive the robot as more extraverted than robots with shorter and scarcer eye contact with a human. In addition to the use of non-verbal behaviors of robots, such as gazes, scholars have used the combination of verbal and non-verbal cues to express extraversion/introversion. Windhouwer (2012) also used non-verbal behaviors of robots, where the robots showed more voice speech and body movement when they were set to be extraverted as opposed to introverted. On the other hand, Aly and Tapus (2013) employed both verbal and non-verbal cues to implement extraversion by making a robot more verbally responsive and bodily reactive to individuals. Also, as Celiktutan and Gunes (2015) reported, the robot's extraversion was manipulated with more frequent hand gestures, faster speech rate, and a higher volume of the robot voice.

The measurement of extraversion and introversion has been done in different ways, including Big Five personality traits (John, Donahue, and Kentle 1991; John and Srivastava 1999), International Personality Item Pool (Wiggins 1979), Myers–Briggs Type Indicator (MBTI; Murray 1990), and Eysenck Personality Inventory (EPI; Eysenck and Eysenck 1987). Although each study adopted slightly different indices to capture human extraversion, the commonality across the studies includes items regarding energy, assertiveness, sociability, and talkativeness in one's behavior. This is a reason the robots were manipulated to have more speech, higher volume, and faster more frequent gestures to demonstrate high extraversion in studies on this topic (e.g., de Graaf and Ben Allouch 2014; Salam et al. 2017). See Table 5.2 for a summary of robot personality inputs.

Table 5.2 Robot Personality Inputs

| Article | Personality Traits | Measure | Manipulation |
|---|---|---|---|
| Celiktutan and Gunes 2015 | Extraversion and introversion | | Altering robot hand gestures, speech rate, and volume for higher levels of extraversion |
| Cruz-Maya and Tapus | Extraversion and | Big Five: Goldberg, 1990 | Survey items |

| 2017 | introversion | | |
|------|--------------|---|---|
| So et al. 2008 | Extraversion and introversion; thinking and feeling | Participant assessment of 8 adjectives (extravert, introvert, rational, kind, aggressive, friendly, strict, and mild) | |
| Lee et al. 2006 | Extraversion and introversion | | Verbal and non-verbal cues |
| Tapus and Matarić 2008 | Extraversion and introversion | Participant assessment of whether a robot has similar personalities with the participant | Verbal and non-verbal cues |
| Joosse et al. 2013 | Extraversion and introversion | | Verbal and non-verbal cues |
| Aly and Tapus 2013 | Extraversion and introversion | | Verbal and non-verbal cues |
| Groom et al. 2009 | Friendliness, integrity, and malice | Friendliness (cheerful, enthusiastic, extraverted, happy, helpful, kind, likeable, outgoing, and warm) Integrity (helpful, honest, pretenseless, reliable, and trustworthy) Malice (disobedient, dishonest, unkind, harsh, and incompetent) | |
| Salam et al. 2017 | Extraversion and introversion | | Robot behaviors |
| Andrist et al. 2015 | Extraversion vs. introversion | | Robot gaze |
| De Graaf and Ben Allouch 2014 | All Big Five personality traits | All Big Five traits/Big Five Inventory (BFI) | |
| Mileounis et al. 2015 | Dominance and extraversion | | Robot behaviors |
| Niculescu et al. 2013 | Extraversion and introversion | | Robot behaviors |
| Dang and Tapus 2015 | Extraversion and introversion | | Robot behaviors with coaching style |
| Aly and Tapus 2016 | Extraversion and introversion | | Robot verbal and non-verbal responsiveness |
| Windhouwer 2012 | Extraversion and introversion | | Robot body movement and voice |

Although extraversion/introversion has been the most common personality trait in the literature of the human–robot personality match, scholars have examined other personality traits. Dominance is one example. Mileounis and colleagues (2015) measured individuals' dominance along with their extraversion and manipulated robots in their experiment to be either domineering or not. Although dominance is not one of the Big Five traits, it is found to be highly associated with extraversion (Digman 1990). Besides, So and colleagues (2008) examined the human–robot personality match in a matrix of four categories: extraversion and thinking, extraversion and feeling, introversion and thinking, and introversion and feeling. In their study, they programmed a feeling robot to say more emotional words and change emotional states more frequently than a thinking robot.

Overall, scholars in HRI acknowledge the importance of the human–robot personality match. Similarities and difference in personality between human and robot have been viewed mostly as binary: same versus opposite in a personality trait. Human personalities were measured using psychometric scales, such as Big Five personality traits and compared with robot personalities in the trait. Extraversion/introversion was the most commonly studied personality trait. In most cases, extravert-type robots were shown to have more lively speech and animated gestures than introvert-type robots. In addition to extraversion, scholars have examined the personality match in dominance, thinking versus feeling, and self-extension. Table 5.3 lists the research on similarities and differences in human and robot personality inputs.

Table 5.3 Human and Robot Similarities/Differences Inputs

| Article | Matched Personalities | Observed Match Type |
|---|---|---|
| Celiktutan and Gunes 2015 | Extraversion and introversion | Similarity |
| Cruz-Maya and Tapus, 2017 | Extraversion and introversion | Similarity & Complementary/Different |
| So et al. 2008 | Extraversion and introversion Thinking and feeling | Similarity |
| Lee et al. 2006 | Extraversion and introversion | Complementary/Different |
| Tapus and Matarić 2008 | Extraversion and introversion | Complementary/Different |
| Joosse et al. 2013 | Extraversion and introversion | Complementary/Different |
| Aly and Tapus 2013 | Extraversion and introversion | Similarity |
| Groom et al. 2009 | Not as an independent variable. Participant personalities were measured as a control variable | Similarity |
| Salam et al. 2017 | Extraversion and introversion | Similarity |
| Andrist et al. 2015 | Extraversion vs introversion, and motivation | Similarity |
| De Graaf and Ben Allouch 2014 | All Big Five personality traits | Similarity |
| Mileounis et al. 2015 | Dominance and extraversion | Similarity |
| Niculescu et al. 2013 | Extraversion and introversion | Similarity |
| Dang and Tapus 2015 | Extraversion and introversion | Similarity |
| Aly and Tapus 2016 | Extraversion and introversion | Similarity |
| Windhouwer 2012 | Extraversion and introversion | Similarity |

## Section 5.2. Outcomes

What happens when a robot and a person have a similar personality? What are the impacts of the human–robot personality match on people's interaction with robots? Scholars in HRI have sought to understand the consequences of similarities and differences of personalities between a human and a robot. In a nutshell, research generally shows that the human–robot personality match leads to positive outcomes in several aspects of human–robot interaction.

To understand the impacts of the human–robot personality match, scholars in HRI have examined several dependent variables. The dependent variables can be categorized into three areas in large part: perceptions of the robot, quality of interaction with the robot, and the likelihood of interacting with the robot in the future. In the following paragraphs we discuss each of these areas.

First, the human–robot personality match is generally found to increase the positive perception of robots by individuals. The positive perceptions are associated with a robot's quality, personality, and competency; these include enjoyment, empathy, intelligence, social attraction, credibility and trust, perceived performance, and compliance. For instance, Lee et al. (2006) tested the intelligence and social attraction of a robotic pet. They also examined social presence, the degree to which a robot elicits social interaction by the interaction as a communication counterpart. Joosse et al. (2013) explored how the human–robot personality match alters human perceptions of trust in the robot, credibility as a source of information, likeability, and intelligence. Mileounis and Barakova (2015) captured the degree to which a robot was perceived as socially intelligible, likable, and emotionally expressive. Windhouwer (2012) examined similar dependent variables with other studies, such as a robot's intelligence, enjoyability, and entertainment. Outcomes in the literature are detailed in Table 5.4.

Table 5.4 Human and Robot Similarities/Differences Outcomes:

| Category | Construct | Paper | Measure |
|---|---|---|---|
| Perception of robot quality | Intelligence | Lee et al. 2006 | A custom scale with two adjective items: competent and clever |
| | | Joosse et al. 2013 | Subset of the Godspeed questionnaire Bartneck et al. 2009 |
| | | Mileounis and Barakova 2015 | Godspeed questionnaire Bartneck et al. 2009 |
| | | Windhouwer 2012 | A custom scale with two adjective items: intelligent and clever |
| | Social attraction | Lee et al. 2016 | An index of three custom items: (a) I think I could spend a good time with this AIBO, (b) I could establish a personal relationship with this AIBO, (c) I would like to spend more time with this AIBO |
| | | Joosse et al. 2013 | Interpersonal attraction scale McCroskey and McCain 1974 |
| | | Mileounis and Barakova 2015 | Godspeed questionnaire Bartneck et al. 2009 |

| | Trust | Joosse et al. 2013 | Source Credibility Scale McCroskey 1973 |
|---|---|---|---|
| | Emotionally expressive | Mileounis and Barakova 2015 | Godspeed questionnaire Bartneck et al. 2009 |
| | Entertaining | Windhouwer 2012 | An index of four personality adjectives: enjoyable, exciting, fun, and satisfying |
| Perception of robot personality | Extraversion and introversion | Andrist et al. 2015 | Not specified |
| | | Lee et al. 2006 | IPIP scale Wiggins 1979 |
| | | Niculescu et al. 2013 | Kahn and De Angeli 2009 |
| | | Joose et al. 2013 | IPIP scale Wiggins 1979 |
| | | Celiktutan and Gunes 2015 | A single custom item: "I thought the robot was assertive and social." |
| | Perception of personality similarity | de Graaf and Ben Allouch 2014 | Saucier 1994 |
| | Self-extension | Groom et al. 2009 | The absolute value of the difference between the participants' rating of themselves and the robot on each trait Kiesler and Kiesler 2005 |
| Quality of interaction with robot | Realism | Celiktutan and Gunes 2015 | A single custom item: "I found the robot behavior realistic." |
| | Enjoyment of interaction | Lee et al. 2006 | An index of three adjective items: enjoyable, fun, and entertaining |
| | Engagement | Salam et al. 2017 | Temple Presence Inventory Lombard et al. 2011 |
| | Stress | Dang and Tapus 2015 | Heart rate signal |
| | Interaction Preference | Cruz-Maya and Tapus, 2017 | Speed; Amplitude; Distance |
| Acceptance | Preference for the robot with similar personality | So et al. 2008 | A custom scale |
| | | Tapus and Mataric 2008 | A single custom item: "The robot personality is a lot like me." |
| | | Dang and Tapus 2015 | A single custom item: "Did you prefer the game with the robot or without the robot?" |

Along with the robot's quality, researchers examined a robot's perceived personality as a dependent variable. For instance, several studies used whether or not people correctly identified a robot's personality, such as extraversion versus introversion, as dependent variables (e.g., Andrist et al. 2015, Lee et al. 2006, Niculescu et al. 2013). Moreover, some studies tested whether people in experiments perceived the similarity in their personality and the robot's (e.g., Aly and Tapus 2013, de Graaf and Ben Allouch 2014). In a similar vein, Groom et al. (2009) used the perception of self-extension to a robot as a dependent variable to capture the similarity between a human and a robot. Self-extension refers to the degree to which an individual considers a robot as part of their identity and projects the identity of himself or herself onto the robot (Connell and Schau 2013).

The second area pertains to the quality of the interaction with the robot. One of the common dependent variables is the enjoyment of the interaction with a robot. Celiktutan and Gunes (2015) measured the realism of the interaction with the robot and enjoyment. Lee and colleagues (2006) also measured enjoyment of the interaction with a robotic pet, which was distinguished from robot perceptions such as social attraction of the robot. Another construct related to the interaction quality is engagement. Engagement captures to what degree individuals are committed and focused on the interaction with a robot. Salam et al. (2017) captured individual and group engagement with the robot by measuring involvement, interest, and enjoyment during the interaction with a robot.

Finally, scholars have examined the impacts of the human–robot personality match on people's willingness to interact with a robot (i.e. acceptance). The construct is captured mainly with the preference for the robot that an individual encountered or interacted with during a study. It is generally understood that high levels of preference for a robot can lead individuals to be more willing to interact with and adopt the robot in the future. Several studies, including So et al. (2008), Tapus and Mataric (2008), and Aly and Tapus (2016), measured participants' preference for a robot that expressed a similar personality with themselves. Dang and Tapus (2015) used preference to play with the robot and the robot's personality.

## Section 5.3. Study Method, Sample, Context & Robot Type

Experiments have been the common methodology to study the human–robot personality match. All fifteen studies identified for the topic of human–robot personality match were conducted with experiments. The most control type was wizard of oz (5), followed by autonomous (3) and pre-defined programed scripts (3). An example of wizard of oz was Celiktutan and Gunes (2015) designed an experiment where a robot asked participants questions for 15 minutes. Groom at al. (2009) used the WoZ method for participants to play a board game with a robotic partner. Mileounis et al. (2015) conducted an experiment employing the WoZ method, where participants played the "Who Wants to be a Millionaire?" game. An example of a study which employed an autonomous approach was Andrist and colleagues (2015). They had participants interact with an autonomous robot to solve the Tower of Hanoi puzzle in their experiment. Finally, an example of a pre-defined programed script was Andrist et al. (2015). They showed a video of a robot interacting with a human and then asked viewing participants to answer a questionnaire.

Sample sizes of experiments varied between 18 and 86 individuals. Most studies were done at the individual level, except that by Salam et al. (2017), who examined the group level as well as the individual level. Most studies in the topic recruited participants from university pools that generally consist of students and staff members. This is in large part the reason ages of participants are mostly younger than 30 years. For example, the majority of participants in Windhouwer (2012) were 18–25 years old, except a few (9.4%) people older than 25 years. In terms of gender, based on the available data, male participants (265) outnumbered female (163) participants. Studies in the topic of human–robot personality match have been conducted using sample populations from: U.S. (3), Netherlands (3), France (3), Korea (1), Singapore (1), German (1), Mexico (1), China (1), Romania (1) and Tunisia (1).

The experiments employed various contexts to study the human–robot personality match. Health care (3) and organizational work setting (3) were the two popular setting followed by home settings (1). For example, Tapus and Mataric (2008) conducted the experiment in a health care setting to explore the role of the personality matching when a robot is a caregiver in post-stroke rehabilitation therapy sessions. Dang and Tapus (2015) and Andrist et al. (2015) employed a similar health care setting where a robot had a caregiver role. Service and work settings also frequently appeared. The context in Aly and Tapus (2013, 2016) was a service encounter where a robot gave advice on restaurants to participants. Joosse and colleagues (2013) employed both home and work settings, where a robot performed cleaning tasks at home and worked as a tour guide, respectively. The Nao robot was the most used type of robot (7), followed by Peopolebot (2), Meka (1), AIBO (1) and the i-robot. At least 10 studies used the term humanoid robot when describing the robot they used while on 2 studies used the word pet to describe their robot.

## Section 5.4. Findings

This section incorporates the variables discussed and presents the findings from the studies on the topic of human–robot personality match. In general, research shows that the human–robot personality match was generally shown to enhance the quality of interaction with a robot, promote positive perceptions of a robot, and predict the higher levels of preference for a robot. However, some studies demonstrated different results from the general findings. The rest of this section introduces the findings of these studies.

One trend in findings across most of the studies is that extravert robots were perceived more positively. For instance, Celiktutan and Gunes (2015) reported that the positive link between human personality and the interaction quality variables was found when a robot was viewed as extraverted. Furthermore, So and colleagues (2008) observed similar findings: when a service robot was viewed as feeling, rather than thinking, people preferred extravert robots to introvert robots. Joosse et al. (2013) also reported that extravert-type robots were trusted and perceived as more credible than introvert-type robots. Windhouwer (2012) demonstrated similar findings, such that individuals found an extravert-type robot was more fun to interact with.

Another trend is associated with the impacts of matching personalities between a human and a robot based on preference. Aly and Tapus (2013, 2016) showed that participants in their experiment preferred the robot that adapted to their personality. Such findings were consistently observed in several studies. Niculescu et al. (2013) showed that introverted people preferred interacting with introvert-type robots. Salam et al. (2017) also demonstrated that generally matching the user and robot personality led to higher levels of engagement with the robot.

The positive relationship was found to be contingent on a robot's characteristics. For instance, So and colleagues (2008) also showed evidence of the personality match for robot preferences:

people with feeling, rather than thinking, preferred feeling robots to thinking robots. Joosse et al. (2013) provided similar evidence that extraverted people liked a robot with extraverted personality characteristics more when the robot was a tour guide, while introverted people liked a robot with the opposite personality when the robot was a cleaner. As found in de Graaf and Ben Allouch (2014), individuals with high expectations of a robot attributed a similar personality to the robot.

However, such positive impacts of matching personalities were not significant in some studies. For instance, personality matching was not found to be significantly associated with positive robot perceptions in Mileounis et al. (2015). Dang and Tapus (2015) also reported no significant impacts of human–robot personality matching. It may be in part because it is not always possible for people to notice different behaviors of robots based on programmed personalities (Andrist et al., 2015). Table 5.5 for a synopsis of predictors and outcomes in previous studies on human and robot personality similarities and differences.

Table 5.5 Human and Robot Similarities/Differences: Predictors and Outcomes

| Predictors | Outcomes |
| --- | --- |
| **Extraversion/introversion:** Aly and Tapus 2013, Andrist et al. 2015, Celiktutan and Gunes 2015, de Graaf and Ben Allouch 2014, Niculescu et al. 2013, Salam et al. 2017, Windhouwer 2012<br><br>**Dominance:** Mileounis et al. 2015<br><br>**Thinking/feeling:** So et al. 2008 | **Perception of robot quality**: Joosse et al. 2013, Lee et al. 2006, Mileounis and Barakova 2015, Windhouwer 2012<br><br>**Perception of robot personality:** Aly and Tapus 2013, Andrist et al. 2015, de Graaf and Ben Allouch 2014, Groom et al. 2009, Lee et al. 2006, Niculescu et al. 2013<br><br>**Quality of interaction with robot:** Celiktutan and Gunes 2015, Lee et al. 2006, Salam et al. 2017<br><br>**Acceptance:** Aly and Tapus 2016, So et al. 2008, Tapus 2015, Tapus and Mataric 2008 |

## Section 5.5. Chapter Summary

In summary, chapter 5 reviews the literature on Thrust Area 3: Robot and Human Personality Similarities and Differences. This literature examines the impact of matching or mis-matching human and robot similarity and/or differences on human robot interactions. This matching or mis-matching was primarily via extraversion and introversion traits. The result were not always consistent regarding whether matching or mis-matching was better. Nonetheless, many studies did find that matching improved the human's enjoyment, empathy, intelligence, social attraction, credibility and trust, perceived performance, and compliance.

# Chapter 6

# Thrust Area 4:

# Factors Impacting Robot Personality

## Section 6.1. Inputs

Studies that have investigated the impact of robot personality have largely utilized five types of independent variables. These independent variables have been (a) robot's behavior(s), (b) robot's physical appearance, (c) robot's role, (d) robot's embodiment and (e) the personality of the human.

***Robot behaviors***. This was the most common independent variable used to invoke robot personality. The behavior studied varies significantly across the literature base and typically depended on the robot's design. For example, in the case of J. Kim, Kwak, and Kim (2009), the robot utilized was a small cylindrical robot (Sony Rolly) that had limited behavioral options while Andrist, Mutlu, and Tapus (2015) used a much more complex robot (Google Meka) with significantly more behavior options. In these studies, behavior comprised two different behavioral elements: physical, and non-physical or communicative behaviors.

First, physical behaviors typically took the form of gestures (Birnbaum et al. 2016; De Ruyter et al. 2005; Hoffman, Birnbaum, Vanunu, Sass, and Reis 2014; Johal, Pesty, and Calvary 2014; H. Kim, Kwak, and Kim 2008; Meerbeek et al. 2008; Moshkina and Arkin 2005; Tay, Jung, and Park 2014; Weiss, Van Dijk, and Evers 2012), movement patterns (Cauchard, Zhai, Spadafora, and Landay 2016; Hendriks et al. 2011; J. Kim et al. 2009; Moshkina and Arkin 2005; Weiss et al. 2012; Woods, Dautenhahn, Kaouri, Te Boekhorst, and Koay 2005; Woods et al. 2007), facial expressions (De Ruyter et al. 2005; Ludewig, Döring, and Exner 2012; Meerbeek et al. 2008), and gaze (Andrist et al. 2015).

Second, communicative behavior was also used as an independent variable. These behaviors typically took the form of audio style (e.g., tone, pitch, volume; Hendriks et al. 2011; Johal et al. 2014; Ludewig et al. 2012; Meerbeek et al. 2008; Sundar et al. 2017; Weiss et al. 2012), written text (Birnbaum et al. 2016; Hoffman et al. 2014), linguistic style (De Ruyter et al. 2005, Meerbeek et al. 2008, Ullrich 2017, Woods et al. 2005, Woods et al. 2007), voice gender (Tay et al. 2014), voice speed (Johal et al. 2014; Tay et al. 2014), and responsiveness (De Ruyter et al. 2005).

***Robot Appearance***. Beyond behavior, robots' physical appearances were also employed as independent variables. The physical appearances largely depended on the robot used in the study. Most robots were humanoid and had faces in either a simulated physical manner (such as the Affetto robot or nurse-bot Pearl) or via a screen attached to the robotic assembly. A notable difference between studies was the physical size of the robot used. Sizes ranged significantly from 58 cm (Nao robot) to about 124 cm (PeopleBot).

***Robot Role***. Robots' role was used as an independent variable in this field of research but was limited to one study. Weiss et al. (2012) investigated the ways that different roles or tasks assigned to a robot might change the perceptions of that robot's personality. The roles in this study were teaching (teacher), convincing (CEO), and caring (pharmacist) and were supported by the environment the robot was seen as operating in. For example, in the caring role, the robot was behind a pharmacy counter (Weiss et al. 2012).

***Robot Embodiment***. Robot's embodiment varied between a physically embodied robot and a disembodied robot on a screen. Most studies utilized a physically present robot (or images of a physically present robot) but, four studies utilized and compared both (Hwang, Park, and Hwang 2013; Kiesler, Powers, Fussell, and Torrey 2008; Ogawa et al. 2009; Walters et al. 2008). All four of these studies utilized a physically embodied robot and a video alternative with the exception of Hwang, 2013 who utilized still images instead of video.

***Human Personality Traits***. Last, human personality traits were also used as an independent variable. These personality traits were gathered with a range of different scales. The personality traits recorded were a participant's levels of extraversion vs. introversion (Andrist et al. 2015; Kimoto et al. 2016; Ogawa et al. 2009; Sandoval et al. 2016; Weiss et al. 2012; Woods et al. 2005; Woods et al. 2007); neuroticism (Kimoto et al. 2016; Sandoval et al. 2016; Woods et al. 2005; Woods et al. 2007); agreeableness, conscientiousness and openness (Kimoto et al. 2016; Ogawa et al. 2009; Sandoval et al. 2016); perceived enjoyment, intelligence, fun, trust, compliance, and willingness to spend time with the robot (Weiss et al. 2012); psychoticism and autonomy (Woods et al. 2005; Woods et al. 2007); motivation (Andrist et al. 2015); and positive, negative, or neutral personality indicators (Ullrich 2017). The measures used for gathering these personality traits were equally as various. A popular commonality among instruments was use of subcomponents of the Big Five personality model. A more detailed presentation of personality traits and measures can be seen in Table 6.1.

Table 6.1 Factors Impacting Robot Personality Inputs

| Article | Personality Traits | Measure |
|---|---|---|
| Sandoval et al. 2016 | Extraversion, agreeableness, conscientiousness, neuroticism or emotional stability, and openness | Big Five traits TIPI Gosling et al. 2003 |
| Andrist et al. 2015 | Extraversion vs. introversion, and motivation | Big Five John and Srivastava 1999<br><br>Intrinsic and extrinsic motivation Guay et al. 2003 |
| Kimoto et al. 2016 | Extraversion, agreeableness, conscientiousness, neuroticism or emotional stability, and openness | Big Five traits John and Srivastava 1999 |
| Weiss et al. 2012 | Extraversion vs. introversion, perceived enjoyment, intelligence, fun, trust, compliance, and willingness to spend time with the robot | Wiggins personality test |
| Woods et al. 2005 | Introversion vs. extraversion, neuroticism, psychoticism, autonomy | Eysenck's Three-Factor Psychoticism, Extraversion and Neuroticism (PEN) model |

| Woods et al. 2007 | Introversion vs. extraversion, neuroticism, psychoticism, autonomy | Eysenck's Three-Factor Psychoticism, Extraversion and Neuroticism (PEN) model |
|---|---|---|
| Ullrich 2017 | Positive, negative, or neutral personality indicators | New survey for new traits |
| Ogawa et al. 2009 | Extraversion, openness, and agreeableness | NEO-FFI |

## Section 6.2. Outcomes

Human Perceptions of Robot Personality. When people interact with robots, they have impressions of the robots in terms of perceived robot personality. Prior literature used different personality questionnaires comprising multiple dimensions of personality. An example is the Big Five personality index, which consists of extraversion, agreeableness, conscientiousness, neuroticism, and openness (Chee et al. 2012; Hendriks et al. 2011; Hwang et al. 2013; Kimoto et al. 2016; Meerbeek et al. 2008; Moshkina and Arkin 2005; Sandoval et al. 2016; Walters et al. 2008). In addition to using the full set of dimensions, some studies picked a subset of personality dimensions of the Big Five index. For example, Ogawa et al. (2009) used the Japanese Property-based Adjective Measurement questionnaire to examine people's perceptions of robot personality on the dimensions of extraversion, openness, and agreeableness. Andrist et al. (2015) and Ludewig et al. (2012) used only the extraversion dimension in their studies.

Aside from the Big Five index, several studies have used alternative measures. For example: sociability, competence, attractiveness, dominance; friendliness; being exhausted, anti-social, an adventurer, etc. (Birnbaum et al. 2016; Cauchard et al. 2016; Groom et al. 2009; Hoffman et al. 2014; Johal et al. 2014; Kiesler et al. 2008; H. Kim et al. 2008; J. Kim et al. 2009; Powers and Kiesler 2006; Ullrich 2017; Walters et al. 2011). It is also worth mentioning that Eysenck's PEN model (1991) was also used in two papers to identify and measure perceived robot's personality dimensions including psychoticism, extraversion, and neuroticism (PEN) (Woods et al. 2005; Woods et al. 2007). In addition, another set of measured different personality traits using various personality questionnaires. Yamashita et al. (2016) used the personality impression questionnaire (PIQ), which involved 46 items that investigated human perceptions of robot personality, while Broadbent et al. (2013) used Asch's checklist of characteristics with 18 personality pairs adapted from Asch (1946), Broadbent et al. (2013), and Yamashita et al. (2016). A more detailed breakdown of the personality dimensions examined and their measures can be seen in Table 6.2.

Table 6.2 Human Perceptions of Robot Personality Outcomes

| Personality | Paper | Personality Dimension | Measure of Personality |
|---|---|---|---|
| Big Five | Sandoval et al. 2016 | Extraversion, agreeableness, conscientiousness, neuroticism/emotional stability, openness | Big Five using TIPI Test Gosling et al. 2003 |
| | Hwang et al. 2013 | | |
| | Hendriks et al. 2011 | | |
| | Walters et al. 2008 | | Big Five Domain Scale from the International Personality Item Pool (IPIP) Goldberg 1999 |
| | Chee et al. 2012 | | |
| | Meerbeek et al. 2008 | | Boeree 2004 |
| | Moshkina et al. 2005 | | Goldberg's Unipolar Big-Five Markers Saucier 1994 |
| | Kimoto et al. 2016 | | Big Five personality traits John and Srivastava 1999 |
| | Ogawa et al. 2009 | Extraversion, openness, agreeableness | Japanese property-based adjective measurement questionnaire Hayashi 1978 |
| | Andrist et al. 2015 | Extraversion | John and Srivastava 1999 |
| | Ludewig et al. 2012 | | BFI-K (Big Five Inventory-Short Version) Rammstedt and John 2005) |
| | Weiss et al. 2012 | | Wiggins 1979 |
| | Tay et al. 2014 | | |
| | Groom et al. 2009 | Friendliness, integrity, malice | |
| (PEN) model | Woods et al. 2005 | Psychoticism, Extraversion and Neuroticism (PEN) | Eysenck 1991 |
| | Woods et al. 2007 | | |
| Various non-Big Five | Birnbaum et al. 2016 | Sociability, competence, attractiveness | Hoffman and Vanunu 2013 |
| | Hoffman et al. 2014 | | |
| | Cauchard et al. 2016 | Exhausted, anti-social, and adventurous | Custom made |
| | Kiesler et al. 2008 | Dominant, trustworthy, sociable, responsive, competent, respectful | |
| | Ullrich 2017 | Positive, neutral, negative | |
| | Walters et al. 2011 | Intelligent, predictable, consistent, fast, polite, friendly, | |

| | | obedient, interesting, attentive | |
|---|---|---|---|
| | J. Kim et al. 2009 | Dominance, friendliness | Ball and Breese 2000 |
| | Powers and Kiesler 2006 | Sociability, knowledge, dominance, human-likeness, masculinity, machine-likeness | Bem 1976 |
| | Johal et al. 2014 | Pleasantness (P), the arousal (A) and the dominance (D) | PAD scale Russell and Mehrabian 1977 |
| Myers–Briggs Type Indicator (MBTI ) dichoto-mies | Kim et al. 2008 | Extraversion–introversion and thinking–feeling | Fong, Nourbakhsh and Dautenhahn 2003 |
| Personality items/pairs | Yamashita et al. 2016 | Reliable, pleasant, calm, etc. | Personality Impression Questionnaire (PIQ) |
| | Broadbent et al. 2013 | Trustworthy, amiable etc. | Asch's checklist of characteristics Asch 1946 |

***Human Attitude toward Robot***. In addition to perceptions of personality, studies have also looked at different attitudes that individuals have toward robots. Overall there eight attitudes were investigated. These attitudes were perceived usefulness, perceived ease of use, perceptions of control, enjoyment, performance, attachment, satisfaction, and animacy. First, perceived usefulness was investigated by Andrist et al. (2015), Meerbeek et al. (2008), and Walters et al. (2011). These studies measured perceptions of usefulness via three scales. Walters and Andrist used custom scales and Meerbeek used an adapted perceived usefulness scale from Van der Heijden (Andrist et al. 2015, Meerbeek et al. 2008, Van der Heijden 2004, Walters et al. 2011).

Second, in addition to perceived usefulness, perceived ease of use was also investigated by Moshkina  and Arkin (2005) and Meerbeek et al. (2008). These studies measured perceived ease of use differently, where Meerbeek used a modified version of Venkatesh and Davis' (2000) ease of use questionnaire and Moshkina used the results of three questions from a custom scale (Meerbeek et al. 2008; Moshkina and Arkin 2005; Venkatesh, Morris, Davis, and Davis 2003).

Third, perception of control was investigated by Meerbeek et al. (2008) via a questionnaire adapted from Hinds et al. (2000). Fourth, enjoyment was investigated by a wider range of authors and was largely measured via questionnaires. Two studies investigating enjoyment produced new items (Moshkina and Arkin 2005; Weiss et al. 2012) and the three remaining used adaptations of scales developed by either Huang et al. in 2001 (Meerbeek et al. 2008), Bartneck et al. in 2008 (Ludewig et al. 2012), or Kanda et al. in 2001 (H. Kim et al. 2008) (Bartneck, Croft, and Kulic 2008; Huang, Lee, Nass, Paik, and Swartz 2001; Kanda, Ishiguro, and Ishida 2001).

Fifth, participants' perceptions of robots' performance were investigated by H. Kim et al. (2008), who used the adjective pairs from Kanda et al. (2001); Hendriks et al. (2011), who used sub-components of the Big Five personality index, and Andrist et al. (2015), who used a custom scale. Sixth, in relation to attachment, Moshkina and Arkin (2005) used four questions from a custom scale (Moshkina and Arkin, 2005). Seventh, satisfaction was investigated by De Ruyter et al. (2005) using an in-house scale developed by De Ruyter and Hollemans (1997). Finally, animacy was investigated by Hendriks et al. (2011) via a think-out-loud qualitative exercise, and by Chee et al. (2012) via a questionnaire adapted from Bartneck et al. (2008). A detailed overview of these attitudes and their measures is detailed in Table 6.3.

Table 6.3 Human Attitude toward Robot Outcomes

| Attitude | Papers | Measure |
|---|---|---|
| Perceived usefulness | Andrist et al. 2015 | Custom scale |
| | Walters et al. 2011 | |
| | Meerbeek et al. 2008 | Van der Heijden 2004 |
| Perceived ease of use | Moshkina et al. 2005 | Custom scale |
| | Meerbeek et al. 2008 | Venkatesh and Davis 2000 |
| Perceived control | Meerbeek et al. 2008 | Hinds 2000 |
| Enjoyment | Moshkina et al. 2005 | Custom scale |
| | Weiss et al. 2012 | |
| | Meerbeek et al. 2008 | Huang et al. 2001 |
| | Ludewig et al. 2012 | Bartneck et al. 2008 |
| | H. Kim et al. 2008 | Kanda et al. 2001 |
| Robot's performance | H. Kim et al. 2008 | Adjective pairs adapted from Kanda et al. 2001 |
| | Hendriks et al. 2011 | Sub-components of the Big Five personality index |
| | Andrist et al. 2015 | Custom scale |
| Attachment | Moshkina et al. 2005 | Custom scale |
| Satisfaction | De Ruyter et al. 2005 | In-house scale developed by De Ruyter et al. 1997 |
| Animacy | Hendriks et al. 2011 | Think-out-loud qualitative exercise |
| | Chee et al. 2012 | Bartneck et al. 2008 |

***Behaviors, Behavioral Intention, and Acceptance***. Behaviors and behavioral intention have been dependent variables in several papers related to robots and personality. The only physical behavior specifically identified as a dependent variable was distance. Nomura et al. (2007) investigated the "allowable distance" of a robot from participants. This was measured via video recording data (Nomura et al. 2007). Preceding actual behavior, behavioral intention was also investigated. Powers and Kiesler (2006) investigated behavioral intention via an online survey containing questions related to whether participants would take a robot's advice. Powers stated that the questionnaire used in this study was based on prior research but did not mention a specific source (Powers and Kiesler 2006). In addition, Meerbeek et al. (2008) used a combination of results of perceived ease of use, perceived usefulness, and enjoyment to investigate intention (in their words "willingness-to-use"). Ludewig et al. (2012) also investigated behavioral intention and did so based on a single survey item (degree of using robot in future). Last, Sundar et al. (2017) adapted three items from Venkatesh et al. (2000) to investigate behavioral intention. Alongside behavioral intention, acceptance has also been investigated: Ludewig et al. (2012) employed indicators of likeability, joy of use, and satisfaction to measure acceptance; in addition, Tay et al. (2014) used three items adapted from Heerink et al. (2010), and De Ruyter et al. (2005) used a modified version of Venkatesh et al.'s (2003) unified theory of acceptance and the use of technology questionnaire.

## Section 6.3 Study Methods, Samples, Contexts and Robot Type

Researchers investigating the impact of robot personality usually used similar study designs. Overall, participants were given a questionnaire, exposed to an experimental condition, and then given a post-questionnaire. The pre-test questionnaires were typically focused on gathering demographic characteristics, but some studies used these questionnaires to identify participants' personality traits or for participants to gain familiarity with different scales used in the study. Studies were on average between-subjects (41.9%) or within-subjects (45.2%), with a small number of mixed between- and within-subject designs making up the remainder.

Most studies did not provide specific information on the status of their participants. Of the studies that did provide specific information, the majority of participants were students or a combination of students and other populations. The remaining were healthy adults, parents, shopping customers or members of a retirement community. In relation to gender, three of the total thirty-one studies failed to provide this information. Of the studies that did report a gender breakdown, most studies had more male participants (738) than female (674) participants. Most studies used fewer than fifty participants (61% of all studies). Ages varied between studies but on average most studies had participants older than 18 years and younger than 35 years. Two exceptions to this assessment were Sundar et al. (2017), with an average age of 80 years, and Ludewig et al. (2012), with an average age of 46 years.

Several studies did not provide geographic information (34.2% of all studies) while the remainder was geographically diverse. Specifically, Japan (4), Germany (3), UK (3), U.S. (3), Netherlands (2), Singapore (2), Korea (1), Israel (1), France (1), New Zealand (1), Sweden 91), China (1), Romania (1), and Tunisia (1). Research on the factors impacting robot personality was conducted mostly in the home setting (1), healthcare setting (7) and organizational work setting (3). The most common interaction control was some type of wizard of oz (10), followed by autonomous (9), and pre-defined scripts (4). However, several studies employed a combination of are one control types. This literature also employed the most diverse set of robots. However, the Nao (4) was the most widely used, followed Peoplebot (3), icat (3), Meka (1), AIBO (1), Lego Mindstrom (1), AMIET (1), Rolly (1), and Affetto (1).

## Section 6.4. Findings

During interactions between humans and robots, both entities have an impact on people's perceptions of the robot. When looking at it from the perspective of a human, the attributes of the individual can influence his or her perceptions of robots. This includes perceived robot personality, attitudes, and acceptance. For example, Woods et al. (2005, 2007) found that participants' gender, age, and technological experience were important in relation to their perceptions of similarity between their personality and the robot's personality. Woods also stated that participants tended to evaluate the robot as being more similar to their personality with respect to the extraversion factor than the factors of psychoticism and neuroticism. In addition, extraverted participants tended to have higher personality ratings for the mechanoid robots and were more likely to adopt anthropomorphic heuristics when interacting with non-human animals and objects (Walters et al. 2008). The importance of user personality was also emphasized in other papers. Ogawa et al. (2009) found that the participants' openness was negatively correlated with the agreeableness and extraversion ratings for the robot. Andrist et al. (2015) found that both extraverts and introverts exhibited significantly greater compliance with the personality-matching robot, and introverts reported a marginal preference for the introverted robot (Andrist et al. 2015). The importance of matching human and robot personalities was also emphasized in two additional studies, by Ullrich (2017) and Kimoto et al. (2016).

From the perspective of the robot, different robot behaviors, appearances, and roles can lead to different perceptions on the part of the human. People can differentiate robots based on different presented personalities and behaviors (Andrist et al. 2015, Cauchard et al. 2016, Hendriks et al. 2011, Hoffman et al. 2014, Kiesler et al. 2008, H. Kim et al. 2008, J. Kim et al. 2009, Meerbeek et al. 2018, Moshkina and Arkin 2005), and evidences show that  extraverted robots are preferred (Meerbeek et al. 2018, Walters et al. 2011) and are perceived to be more likeable, friendly, pleasant and socially acceptable (Ludewig et al. 2012). Also, responsive robots are perceived as more sociable (Kimoto et al. 2016, Meerbeek et al. 2008, Weiss et al. 2012), socially intelligent robots are indeed perceived as more socially intelligent, and dominant robots are perceived as more authoritative (Johal et al. 2014).

Robot appearance was also an important factor that influenced people's perceptions. Humanoid robots were favored by people because of the robot's higher perceived control (Groom et al. 2009, Walters et al. 2008), greater friendliness (Chee et al. 2012), more salient personality traits, relatively low degree of eeriness, and higher degree of trustworthiness (Broadbent et al. 2016). Also, robot size, shape, and texture had effects on people's perception. For example, robots with short chin length were perceived as more sociable (Powers et al. 2006) and robot shape impacted participants' emotions and perceptions of the robot's personality (Hwang et al. 2013). In addition, robots were perceived as having a more likeable personality and less dominant personality when providing natural touch sensation to participants (Yamashita et al. 2016). Robot role was also found to influence people's perceptions of robots they interacted with. Tay et al. (2014) found that people were inclined to interact with the robots whose personalities conformed to the robot's occupational role (Tay et al. 2014). Assistant robots that presented a playful personality were perceived as more socially attractive and intelligent, while companion robots were evaluated as less anxious and less eerie when their personality was serious (Sundar et al. 2017). Table 6.4 illustrates the findings from the studies mentioned and organizes them by predictors and outcomes.

In summary, it is clear from the above set of findings that, generally, extraverted robots are favored by most all participants. People also tend to prefer robots that are more like themselves than not. Additionally, robots appearing as humanoid are preferred as opposed to robots appearing mechanoid. Mechanoid robots are, however, acceptable to people with extraverted personalities but not introverted personalities. People also have a higher willingness to interact with robots that are assigned personalities that match the robot's occupational role. Notably, this body of research does not contradict itself, but many questions remain unanswered.

Table 6.4 Factors Impacting Robot Personality: Predictors and Outcomes

| Personality Predictors | Outcomes |
|---|---|
| **Human** | |
| **Human personality:** Andrist et al. 2015; Kimoto et al. 2016; Ogawa et al. 2009; Sandoval et al. 2016; Weiss et al. 2012; Woods et al. 2005; Woods et al. 2007<br><br>**Human demographic information:** Woods et al. 2005, Woods et al. 2007 | **Robot personality:** Walters et al. 2008; Ogawa et al. 2009; Woods et al. 2005; Woods et al. 2007<br><br>**Attitude:** Andrist et al. 2015; Kimoto et al. 2016; Ullrich et al. 2017<br><br>**Acceptance:** Walters et al. 2008; Andrist et al. 2015 |
| **Personality Predictors** | **Outcomes** |
| **Robot** | |
| **Robot behaviors**: Birnbaum et al. 2016; De Ruyter, Saini, Markopoulos, and Van Breemen, 2005; Hoffman, Birnbaum, Vanunu, Sass, and Reis 2014; Johal, Pesty, and Calvary 2014; H. Kim, Kwak, and Kim 2008; Meerbeek, Hoonhout, Bingley, and Terken 2008; Moshkina and Arkin 2005; Tay, Jung, and Park 2014; Weiss, Van Dijk, and Evers 2012<br><br>**Robot appearance:** Broadbent et al. 2016, Chee et al. 2012, Groom et al. 2009, Hwang et al. 2013, Powers et al. 2006, Walters et al. 2008, Yamashita et al. 2016<br><br>**Robot roles:** Sundar et al. 2017, Tay et al. 2014 | **Robot personality:** Andrist et al. 2015; Broadbent et al. 2016; Cauchard et al., 2016; Hendriks et al. 2011; Hoffman et al. 2014; Hwang et al. 2013; Johal et al. 2014; Kiesler et al. 2008; Kim et al. 2008; Kim et al. 2009; Kimoto et al. 2016; Ludewig et al. 2012; Meerbeek et al. 2018; Moshkina and Arkin 2005; Meerbeek et al. 2008; Powers et al. 2006; Sundar et al. 2017; Weiss et al. 2009; Yamashita et al. 2016<br><br>**Attitude:** Meerbeek et al. 2018, Walters et al. 2011;Groom et al. 2009; Walters et al. 2008<br><br>**Acceptance:** Tay et al. 2014 |

## Section 6.5. Chapter Summary

In summary, chapter 6 reviews the literature on Thrust Area 4: Factors Impacting Robot Personality. This literature focuses on approaches to manipulating human perception of the robot's personality. The most common approach to manipulating human perception of the robot's personality was to alter the robot's behaviors. Physical movement behaviors included gestures, movement patterns, facial expressions and gaze. communicative behavior included audio style  written text, linguistic style, voice gender, voice speed and responsiveness.

# Chapter 7

# Major Findings and A Way Forward

## Section 7.1. Major Findings

We derived four major findings from the literature review, listed next. There is also empirical evidence with regard to other findings, but these insights represent the most consistent and generalizable results.

1. Extraverts seem to respond more favorably when interacting with robots.
2. Varying the robots behavior and vocal cues can invoke an extraverted personality.
3. Humans respond more favorably to extravert-type robots, but this relationship is moderated.
4. Humans respond favorably to robots with similar or different personalities from them.

## Section 7.2. Critique of the Major Findings

***Finding 1.*** According to the articles we reviewed, a number of personality traits can be important. The levels of empirical support found for each personality trait vary considerably. Nonetheless, the literature suggests that extraversion plays a key role in understanding human–robot interactions. Extraverts are more receptive to robots, and humans in general are more open to extraverted robots. There are several possible explanations for the findings related to extraversion. First, extraversion as a human trait is a strong predictor of whether someone will engage with someone else (Peeters et al. 2006). Based on the current literature, this effect seems to translate over to human–robot interactions.

***Finding 2.*** Another explanation is that extraversion as a robot trait is easier to display in robots and might be more salient in shorter interaction times. For example, researchers have investigated such behaviors by manipulating the robot's hand gestures, speech rate and volume, along with its speed and frequency of movement (Celiktutan and Gunes 2015; Cruz-Maya and Tapus 2017). However, it is less clear how to have the robot display behavior that would indicate openness to experiences or many other traits. To do so might require advanced technological approaches that many social science researchers typically do not employ. The current literature has also relied primarily on experimental studies conducted over a short duration of time. The impacts of other more subtle traits might not be salient in such a short time.

***Finding 3.*** The importance of robot extraversion in many studies might also be the result of the social nature of the interactions involved in the studies. Researchers in several studies have found evidence of moderators on the impact of robot personality on human–robot interactions. For example, extraversion was found to be less important when a robot was a security robot than when it was a health care robot (Tay et al. 2014). According to Tay et al. (2014), humans expect health care providers to be more social or outgoing, which is less true for security providers. If more studies had examined less-social-oriented interactions between humans and robots, extraversion might not have emerged as being so important.

***Finding 4.*** Unfortunately, we know little about the influence of moderators on the impacts of human personality on human–robot interactions. The social nature of the task in these studies might also make extraversion more important. For example, humans engaging robots with regard to receiving technical knowledge from the robot might make humans' trait of conscientiousness more important and extraversion less important to determining their trust in the robot. In short, the focus on social interactions might help to explain the importance of extraversion as a human trait.

A small but growing number of studies are focusing on the impact of similar vs. different human and robot personalities. This literature has the potential to reframe the discussion around the importance of both human and robot personalities. Nonetheless, there is still a need to explore the impacts of human and robot personalities separately from similar vs. different personalities. Robots do not always know what particular personality a human has; therefore, it is still important to explore the impact of human and robot personalities separately from this thrust area.

***Limitations.*** This literature review has several limitations. First and foremost, no literature review is ever completely inclusive. In particular, we limited this review to English-speaking journals and articles. In this literature review we did not include studies examining EVA robots.

## Section 7.3. A Way Forward

Despite the importance of personality in human–robot interaction and the efforts of many scholars, there are several major gaps. Next we present research opportunities in the literature based on important gaps. These include research opportunities related to context, method, new tasks and personality traits. We discuss these in greater detail next.

***Research Opportunity 1: Taking Context into Consideration.*** No study examined the effects of context on the impacts of human and robot personality. Context has been shown to be important to understanding many different phenomena of interest across research domains. Home and work settings represent two contexts in the human–robot interaction literature. It is easy to imagine that robot personality might be more or less important for home robots than for robots used at work. Gaps in context are likely to hide important contingency variables needed to better understand the impact of personality on human–robot interaction.

***Research Opportunity 2: Leaving the Lab.*** Gaps in research approaches present a major challenge to the generalizability of the results in the literature. There were four major gaps with regard to research approaches. First, most of the studies took an experimental approach. Second, robots are expected to play a major role in the health care industry, but there is a lack of studies in that context (Broadbent et al. 2009). Third, a related shortcoming is the lack of studies over time. Prior literature has highlighted the influence of appropriation over time in understanding human–technology interaction. Yet no work has been done to understand how the impact of personality might change over time. Fourth, although some studies conducted interviews to supplement or complement quantitative analysis, little effort has been made to employ a qualitative approach as the primary method or analysis. Yet, qualitative approaches provide a unique and rich set of insights.

***Research Opportunity 3: New Tasks***. As stated in findings 3 above, the importance of extroversion may be due to the task type. Taking a step back, it seems reasonable that robot personality itself may be more or less relevant depending upon the nature of the task being performed. Future research should be directed at better understanding the role of task type in understanding the importance of personality in human robot interaction.

***Research Opportunity 4: Beyond the Big Five.*** Most of the studies examined one or more of the Big Five personality traits, with extraversion/introversion being the most popular. However, there are many other types of personality measures. For example, only one study claimed to employ the Myers–Briggs personality test (see Kim et al. 2008). It is not always clear why most studies have focused on the Big Five.

## Section 7.4. Chapter Summary

In summary, chapter 7 presented and discussed major findings, critique of the major findings, and way forward. In particular, extroversion as a trait for either human or robots seem to have a positive association with human robot interactions. The literature on the effects of human robot personality similarity and differences is much less clear. Finally, chapter 7 highlights potential opportunities in the study of personality in human robot interactions research such as: including more context, engaging in more field research and focusing on other non-big five personalities.

# Chapter 8